\documentclass[%
reprint,
superscriptaddress,
preprintnumbers,
nofootinbib,
amsmath,amssymb,
aps,
rmp,
floatfix,
a4paper,
showkeys,
]{revtex4-2}

\usepackage{graphicx}
\usepackage{dcolumn} 
\usepackage{bm} 
\usepackage{blindtext}
\usepackage{color}
\usepackage{hyperref}
\usepackage{siunitx}
\usepackage{booktabs}

\graphicspath{{fig/}}
\hypersetup{
  colorlinks=true, 
  citecolor=blue, 
  linkcolor=blue, 
  urlcolor=blue 
}

\begin{document}


\title{kikuchipy: an open-source toolbox for analysis of EBSD patterns}

\author{Håkon W. Ånes}
\email{hwaanes@gmail.com}
\affiliation{Xnovo Technology ApS, 4600 Køge, Denmark}
\affiliation{Department of Materials Science and Engineering, Norwegian University of Science and Technology, N-7491, Trondheim, Norway}
\author{Phillip Crout}
\affiliation{Department of Materials Science \& Metallurgy, University of Cambridge, 27 Charles Babbage Road, Cambridge, United Kingdom}
\author{Lars Andreas Lervik}
\affiliation{Department of Materials Science and Engineering, Norwegian University of Science and Technology, N-7491, Trondheim, Norway}
\author{Ole Natlandsmyr}
\affiliation{Department of Materials Science and Engineering, Norwegian University of Science and Technology, N-7491, Trondheim, Norway}
\author{Tina Bergh}
\affiliation{Department of Chemical Engineering, Norwegian University of Science and Technology, N-7491, Trondheim, Norway}
\author{Jarle Hjelen}
\affiliation{Department of Materials Science and Engineering, Norwegian University of Science and Technology, N-7491, Trondheim, Norway}
\author{Antonius T. J. van Helvoort}
\affiliation{Department of Physics, Norwegian University of Science and Technology, N-7491, Trondheim, Norway}
\author{Knut Marthinsen}
\email{knut.marthinsen@ntnu.no}
\affiliation{Department of Materials Science and Engineering, Norwegian University of Science and Technology, N-7491, Trondheim, Norway}

\date{\today}

\begin{abstract}

We present kikuchipy, an open-source toolbox for analysis of electron backscatter diffraction patterns, written in Python.
The software is capable of both Hough and dictionary indexing and orientation and/or projection center refinement of patterns stored in file formats from all major vendors.
Indexing results can be validated using maps independent of indexing and by visually comparing experimental and simulated patterns.
By leveraging scientific packages in the Python ecosystem, emphasis is put on making the indexing workflow flexible and improve results through fast iteration.
The software's capabilities are demonstrated on three application examples: analysis of orientation relationships in a super duplex stainless steel, phase differentiation of aluminium and silicon in a cast modified Al--Si alloy, and phase differentiation of particles in an Al--Mn alloy as Al$_6$Mn or $\alpha$--AlMnSi. The diffraction patterns and analysis workflows are made publicly available.
kikuchipy was created and is developed as a resource for the electron microscopy community, allowing anyone to improve the software or include it into their own analysis workflows or softwares.

\end{abstract}


\maketitle


\section*{Introduction}

\begin{figure}[htbp]
  \centering
  \includegraphics[width=0.65\columnwidth]{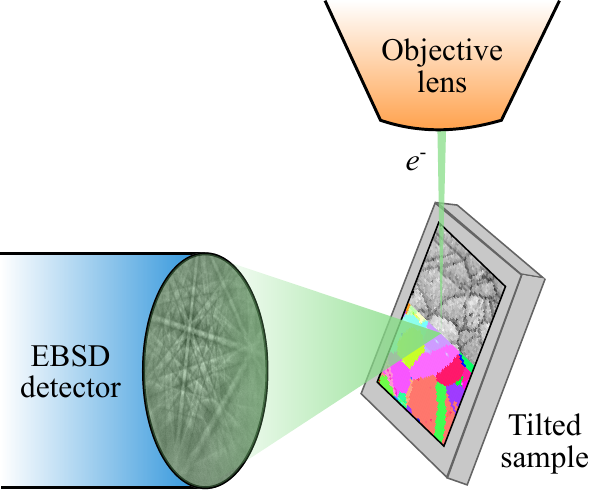}
  \caption{EBSD acquisition within the SEM.}
  \label{fig:ebsd-acquisition}
\end{figure}

Electron backscatter diffraction (EBSD) is a technique to characterize crystallographic features in the scanning electron microscope (SEM) \cite{schwartz2009electron}.
A typical setup for acquisition of an EBSD dataset within the SEM is illustrated in Fig. \ref{fig:ebsd-acquisition}.
An electron beam is rastered in a grid across a sample surface while collecting a two-dimensional (2D) electron backscatter pattern (EBSP) per beam position.
EBSD is most often used to estimate the orientation of the strongest scattering crystal within the interaction volume in each beam position.
This is called indexing and results in rich surface maps.
Based on pioneering work by research groups at Yale \cite{wright1992automatic} and Risø \cite{lassen1992image}, indexing is automated and widely available, typically in commercial solutions.
The active development of indexing with openly available algorithms has been limited, although a few open-source projects exist, notably the EMsoft suite of command-line programs \cite{jackson2019dictionary}, the EMSphinx graphical user interface (GUI) \cite{lenthe2019spherical}, the AstroEBSD MATLAB software \cite{britton2018astro}, and the PyEBSDIndex Python software \cite{rowenhorst2024fast}.
With kikuchipy, we present a widely available and accessible Python toolbox for further development of EBSP analysis.

There are many sources of error when indexing an EBSP \cite{ram2015error,ram2017error}.
Incorrect indexing may occur unexpectedly, and it might be hard to identify why.
Examples of unexpected results are mis-indexing different crystallographic phases as the same phase, or a scatter of single points of highly different orientations within one grain or along a grain boundary.
It is crucial that the indexing steps are transparent and that we are able to apply them in the desired order when troubleshooting problematic cases.
Complete insight is generally unavailable for commercial solutions but is available for the mentioned open-source software.
However, it is our experience that the software forsake flexibility to increase user-friendliness in terms of automation and ease-of-use.
kikuchipy addresses this issue.
The software's primary aim is to make the off-line indexing workflow---from inspecting the acquired patterns to validating the indexing results---flexible and iterable.
Analysis workflows can be built with flexibility since functionality is provided as independent tools.
Indexing steps can then be optimized to identify and eliminate sources of error.
Hough indexing and pattern matching approaches \cite{chen2015dictionary,nolze2016pattern,singh2017application} are available.


kikuchipy has already been used to improve the signal-to-noise ratio (SNR) of EBSPs \cite{aanes2020processing} and to characterize a range of materials: steels \cite{akselsen2021effect,bugten2023role}, a nickel-based superalloy \cite{akselsen2022microstructure}, aluminium (Al) alloyed with manganese (Mn) \cite{aanes2022correlated,aanes2023orientation}, an Al–steel weld \cite{bergh2023intermetallic}, and ceramics \cite{schultheiss2022confinement,sandvik2023pressure}.
As with previous open-source EBSD software \cite{pinard2011open}, kikuchipy is developed as a research and development tool for the EBSD community.
As such, emphasis is put on flexibility rather than high indexing speeds.

\begin{figure}[htbp]
  \centering
  \includegraphics[width=0.6\columnwidth]{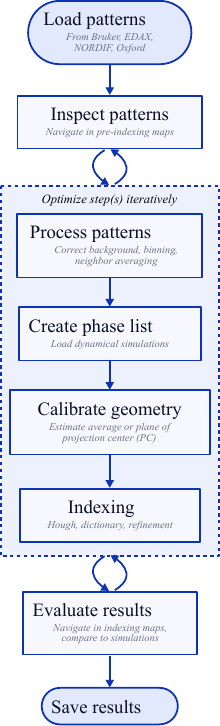}
  \caption{Typical workflow for off-line indexing of a dataset of EBSPs using kikuchipy.}
  \label{fig:workflow}
\end{figure}

This paper showcases the capabilities of kikuchipy.
The focus is on the workflow of indexing an acquired EBSD dataset.
Fig. \ref{fig:workflow} depicts a general analysis workflow using kikuchipy.
The workflow consists of initial steps of loading and inspecting patterns, indexing steps to be optimized, and steps of visual validation and final saving of the indexing results.
We give three examples of applying this workflow to challenging metallurgical datasets using kikuchipy.
Example I demonstrates how to characterize and visualize orientation relationships in a super duplex stainless steel (SDSS) with a substantial fraction of undesirable secondary phases.
Example II addresses the issue of differentiating between similar crystallographic phases using EBSD only by indexing aluminium and silicon in an Al--Si alloy.
Example III concerns phase differentiation of micron-sized second-phase particles in an Al--Mn alloy as Al$_6$Mn or $\alpha$–AlMnSi.
The raw EBSD data and workflow notebooks are made publicly available for verification and reuse from Zenodo at \href{https://doi.org/10.5281/zenodo.20290635}{https://doi.org/10.5281/zenodo.20290635}.

\section*{Software capabilities and workflow}
\label{sec:software}

kikuchipy is implemented in Python and supports Windows, macOS, and Linux.
The software can be installed from the Python package index (PyPI) manager, the Anaconda package manager, or from source at https://github.com/pyxem/kikuchipy.
kikuchipy extends the HyperSpy library \cite{pena2017electron,hyperspy_concept_doi} for EBSP analysis.
HyperSpy is established within the electron microscopy community and provides tools for interactive analysis of multi-dimensional datasets.
kikuchipy follows HyperSpy's object-oriented design, with most analyses available as methods of an EBSD class.
Many file readers are available, enabling file format-agnostic workflows that can be shared with colleagues or alongside a publication.
Datasets of sizes larger than available computer memory can be analyzed in parallel using Dask \cite{rocklin2015dask}.
To reduce computation time, performance-critical parts are compiled with Numba's just-in-time compiler \cite{lam2015numba}.
Two other libraries are essential to kikuchipy's functionality: orix \cite{johnstone2020density,orix_concept_doi} for handling rotations and vectors using crystal symmetry and diffsims \cite{diffsims_concept_doi} for parts of the diffraction simulations.

Users interact with kikuchipy by interactive scripting in a notebook \cite{kluyver2016jupyter} or console or by running a script.
The documentation\footnote{\url{https://kikuchipy.org}.\label{doc}} provides many examples and tutorials for users to build upon when creating workflows suited to their analyses.
The documentation includes a complete description of all functionality and parameters as well.


\subsection*{Loading and inspection of data}
\label{sec:loading-inspection-data}

As shown in Fig. \ref{fig:workflow}, most workflows begin by loading an experimental dataset of EBSPs.
kikuchipy can read files from commercial solutions from Bruker, EDAX, NORDIF, and Oxford Instruments.
Before indexing, we recommend getting an overview of the information in the dataset from so-called `pre-indexing' maps.
The maps are derived solely from pattern intensities and are independent of any user bias introduced in the indexing setup.
They can inform the selection of candidate phases for indexing and help when validating the indexing results.
Three such maps are described and used in the first application example.
Other maps, such as the `quality metrics' maps suggested by \textcite{pinard2011open}, may also be obtained by applying HyperSpy's flexible mapping tools.
We can inspect patterns by navigating any map with the visualization tools from HyperSpy.


\subsection*{Pattern processing}
\label{sec:pattern-processing}

\begin{figure}[htbp]
    \centering
    \includegraphics[width=\columnwidth]{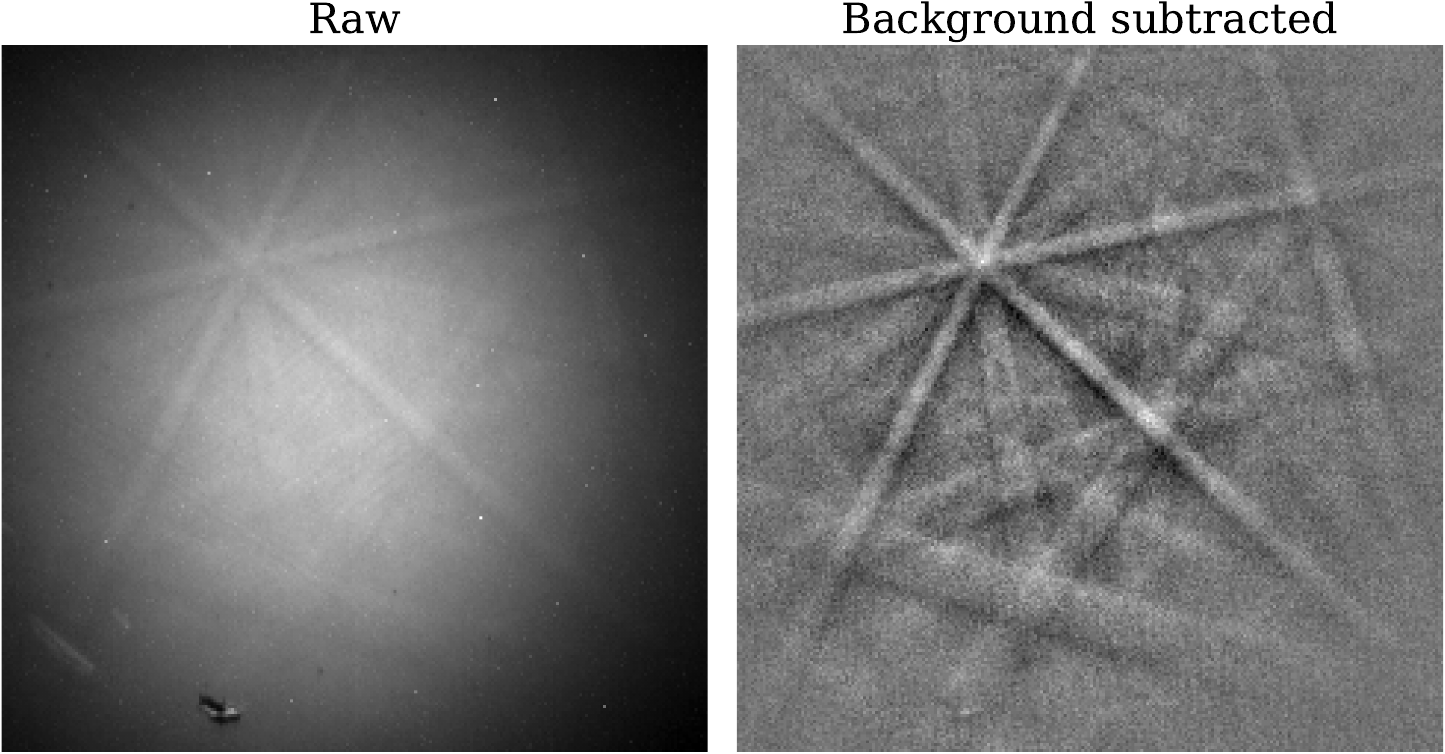}
    \caption{Example of (left) a raw aluminium EBSP as acquired and (right) after static and dynamic background subtraction.}
    \label{fig:pattern-processing}
\end{figure}

Informally, we can interpret an EBSP as the superposition of a diffraction pattern and a smooth background intensity \cite{nolze2017electron}.
An example of a raw EBSP from an aluminium grain is shown in Fig. \ref{fig:pattern-processing}.
Indexing of the diffraction pattern generally requires correction of the smooth background.
Correction is usually carried out in the two steps of static followed by dynamic background correction \cite{wright1992automatic}.
In static background correction, a single background pattern, ideally containing only the smooth background and any constant features on the detector, is removed from every pattern by subtraction or division.
Some detector vendors allow the user to store patterns after static background correction, in which case this step should be skipped.
Static background correction usually leaves some long-range intensity variations in the pattern that can be reduced by dynamic background correction.
In this correction, a dynamic (per pattern) background is obtained by Gaussian blurring and removed by subtraction or division.
The effect of static and dynamic background correction on the raw aluminium EBSP is shown in Fig. \ref{fig:pattern-processing}.

The SNR of patterns in a dataset may be so low that indexing fails \cite{wright2015introduction}.
If this is the case, indexing may be improved by averaging each pattern with its nearest neighbors, e.g. using a $3 \times 3$ Gaussian kernel centered on the pattern to average \cite{aanes2020processing}.
If indexing does not require a high pixel resolution, such as when differentiating between similar phases or obtaining a high orientation accuracy, reducing the number of pattern pixels by binning to increase the indexing speed can be advantageous.
Other processing steps such as adaptive histogram equalization \cite{marquardt2017quantitative}, rescaling, and normalization of pattern intensities are also available.

Writers of EBSPs to select file formats such as the open Hierarchical Data Format Version 5 (HDF5) are available.
These allow users to reload processed patterns with kikuchipy at a later time or share the patterns with others.
Other software such as EMsoft \cite{jackson2019dictionary} can directly read patterns stored in some of these formats.
This interoperability enables users to combine different software in their analysis workflow.
Interoperability is one of the four `FAIR' guiding principles for scientific data management and stewardship (Findable, Accessible, Interoperable, and Reusable) \cite{wilkinson2016fair}.


\subsection*{Geometry calibration}
\label{sec:geometry-calibration}

Indexing an EBSP requires a calibrated detector-sample geometry.
The geometry can be considered as the detector's view of the sample beam position.
The geometry is described by the so-called projection/pattern center (PC), the sample and detector tilts, and a rotation of the detector or sample about the optic axis (usually \SI{0}{\degree}).
The PC is the shortest distance from the beam position on the sample surface to the detector and consists of three parameters (PC$_{\mathrm{x}}$, PC$_{\mathrm{y}}$, PC$_{\mathrm{z}}$).
(PC$_{\mathrm{x}}$, PC$_{\mathrm{y}}$) describe the PC on the 2D detector and PC$_z$ describes the distance to the sample beam position.
These parameters and tilts allow us to systematically relate the EBSP on the detector to the crystal orientation in the sample reference frame.
The relation involves consistent transformations between reference frames for the detector, sample, and crystal.
Commercial solutions and open-source software use different conventions when defining these reference frames.
\textcite{britton2016tutorial} describe the conventions used in AstroEBSD and one commercial solution.
The conventions adopted in kikuchipy are given in the documentation\footref{doc}.

The size of a single detector pixel on EBSD CCD detectors is usually in the range \SIrange{50}{100}{\micro\meter} \cite{jackson2019dictionary,aanes2020processing}.
Using a fixed PC for all patterns in a dataset is a reasonable approximation if the scanned area is comparable to the pixel size \cite{singh2017application}.
However, the PC may vary over several detector pixels for larger scanned areas.
In such cases we need a dynamic PC that changes between map points.
A reasonably good guess for the PC can be optimized by Hough indexing in PyEBSDIndex \cite{rowenhorst2024fast}.
Hough indexing routines in PyEBSDIndex are wrapped in kikuchipy for ease-of-use and closer integration with the other tools available in the EBSD class.
Parameters required by the optimization that are fixed include the sample and detector tilts and a crystal phase.
Further refinement of the PC can be achieved by pattern matching.
We can validate PCs by comparing experimental and simulated patterns.
If we need a dynamic PC, we can either fit a plane to selected PCs \cite{winkelmann2020refined} or extrapolate from a mean PC \cite{singh2017application}.
Extrapolation requires that we know the pixel size of the detector.


\subsection*{Indexing}
\label{sec:indexing}

Two indexing approaches are available in kikuchipy: Hough indexing (via PyEBSDIndex) and dictionary indexing \cite{chen2015dictionary}.
Indexing results can be further improved in a refinement step \cite{nolze2016pattern,singh2017application}.
Both orientation and PC refinement are possible.
Indexing results from other software can also be refined by loading these into Python with e.g. the readers in the orix library (see that package's documentation\footnote{\url{https://orix.readthedocs.io}.} for details).
An a priori known crystal phase is required.

Hough indexing relies on detecting Kikuchi band positions in the Hough transform.
An orientation is estimated by consistent assignment of crystallographic indices to the bands.
\textcite{britton2018astro} and \textcite{rowenhorst2024fast} describe in detail the implementation of Hough indexing in the AstroEBSD and PyEBSDIndex software, respectively.
Hough indexing is generally fast and requires a minimum of three crossing bands to estimate the orientation.
In kikuchipy, Hough indexing is preferred over dictionary indexing for quick estimates of the orientation or PC from an EBSP, such as when calibrating the detector-sample geometry.
However, if the SNR is too low, too few or no bands are detected and thus no estimation is given \cite{wright2015introduction,singh2018high,aanes2020processing}.
For EBSPs where there are bands present but Hough indexing fails, dictionary indexing should be used.

The implementation of dictionary indexing is adapted from EMsoft.
In this indexing method, every EBSP is compared to a dictionary of simulated patterns.
The orientation of the best-matching dictionary pattern is taken as the estimated orientation of the EBSP.
To generate a dictionary, we need (1) a simulated master pattern in the stereographic or square-Lambert projection of a crystal phase, (2) a discrete uniform sampling of orientations within the fundamental zone of a phase's proper point group, and (3) a detector-sample geometry with a fixed PC.
One dictionary per phase in the phase list is required.
Best results are in general obtained using dynamically simulated master patterns \cite{winkelmann2007many,callahan2013dynamical}.
All dynamical simulations in this work are generated with EMsoft.

Dictionary orientations can be obtained by cubochoric sampling of a fundamental zone \cite{singh2016orientation} as implemented in orix.
We can control the average misorientation angle $\left<\omega\right>$ between sampled orientations.
A lower $\left<\omega\right>$ generally gives a better estimation of the orientation of the EBSP at the cost of a lower indexing speed due to a larger dictionary.
\textcite{jackson2019dictionary} suggest that $\left<\omega\right>$ = \SI{1.4}{\degree} is a reasonable trade-off.
Note that the number of sampled orientations increases with a reduction in proper point group symmetry, as this corresponds to a larger fundamental zone.
For example, cubochoric sampling returns about 300 000 orientations for the proper point group $432$ and 600 000 orientations for the proper point group $622$ at $\left<\omega\right> = $ \SI{1.4}{\degree}.
This average misorientation is used for all orientation sampling for dictionary indexing in this work.

Here, as in previous work \cite{nolze2015kikuchi}, we use the normalized cross-correlation (NCC) score between two patterns as the similarity metric to find a best match.
The score is defined as \cite{gonzalez2017digital}

\begin{equation}
  r = \frac{\sum^n_{i=1}(x_i - \bar{x})(y_i - \bar{y})}{\sqrt{\sum ^n _{i=1}(x_i - \bar{x})^2}\sqrt{\sum ^n _{i=1}(y_i - \bar{y})^2}},
  \label{eq:ncc}
\end{equation}

\noindent where experimental patterns $x$ and simulated patterns $y$ are centered by subtracting out the mean of each pattern and the sum of cross-products of the centered patterns is accumulated.
The denominator adjusts the scales of the patterns to have equal units.
The value of $r$ is generally between 0 and 1.
Here, we use this metric in both dictionary indexing and refinement.

The discrete sampling for dictionary indexing means it is unlikely that the true crystal orientation corresponds exactly to a dictionary orientation.
This may be fine for analysis of single-phase materials requiring a rough orientation estimate only.
For other analysis, we recommend to refine the orientations.
In kikuchipy, the user can choose between refining the orientation, the PC, or both at the same time.
The orientation is described by the three Euler angles, meaning that either three or six parameters are refined.
Orientation refinement allows for a fixed PC for all patterns or one PC per pattern.
Refinement is performed by allowing the chosen parameters to vary in a controlled manner while optimizing the NCC score $r$.
Refinement can be exited when the improvement in $r$ between iterations is lower than a given value, such as \SI{1e-4}, or when a given number of iterations is reached, such as 100.
Many optimization algorithms from SciPy \cite{virtanen2020scipy}, both local and global, are supported, as well as the Nelder-Mead simplex from NLopt \cite{nlopt}.
The latter is used in the application examples in this work as it is found to be slightly faster.
For orientation refinement after dictionary indexing in this work, we search for the final best-fit Euler angles within $\pm$\SI{5}{\degree} from the initial estimate given by dictionary indexing.
Also, refinement can be stopped when the improvement in $r$ between iterations is lower than \SI{1e-4}..


\subsection*{Validation of indexing results}
\label{sec:validation-indexing-results}

\begin{figure*}[htbp]
    \centering
    \includegraphics[width=0.75\textwidth]{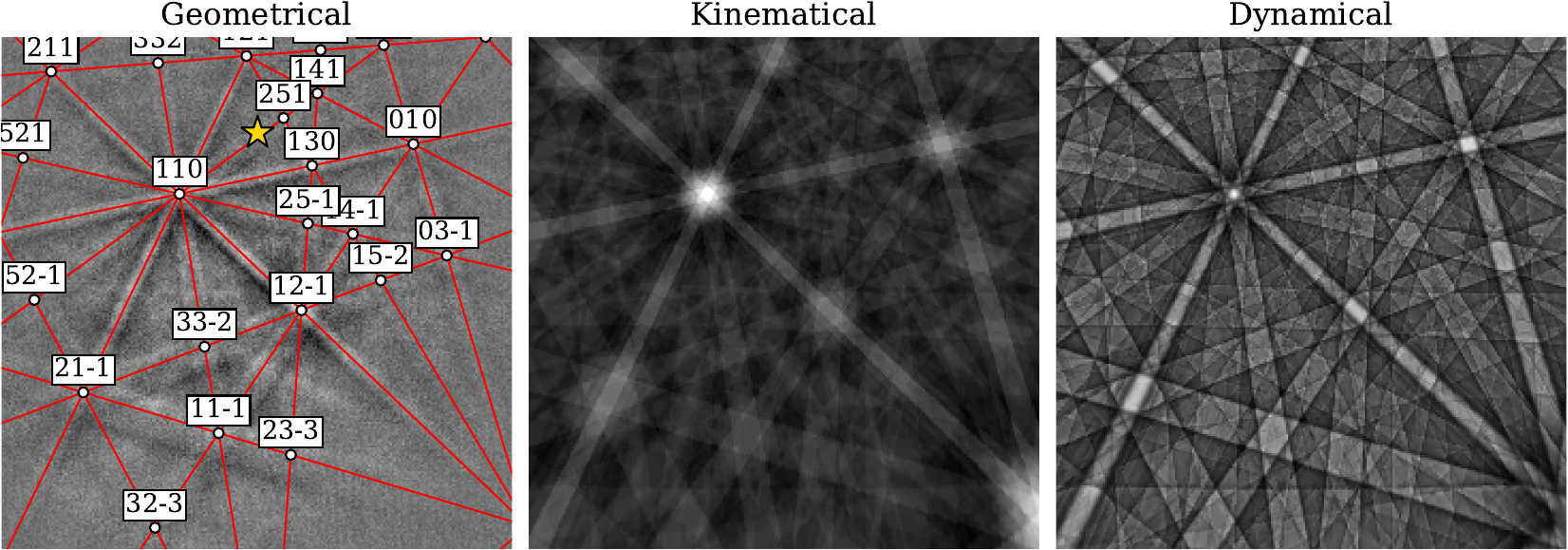}
    \caption{
    Visual validation of indexing results using (left) geometrical, (middle) kinematical, (right) and dynamical simulations for the aluminium EBSP in Fig. \ref{fig:pattern-processing}.
    The geometrical simulation includes bands, corresponding zone axes and their labels, and the PC as a gold star.
    The dynamically simulated EBSP is projected from a master pattern generated with EMsoft.
    }
    \label{fig:pattern-validating-results}
\end{figure*}

Powerful visualization tools in HyperSpy allow us to validate our indexing results by comparing indexed solutions to experimental EBSPs.
In kikuchipy, indexed solutions can be represented by geometrical, kinematical, or dynamical simulations, the latter requiring a dynamically simulated EBSD master pattern.
To demonstrate, the orientation and PC for the aluminium EBSP in Fig. \ref{fig:pattern-processing} is estimated with Hough indexing followed by refinement using dynamical simulations.
Visual validation of the indexing results is presented in Fig. \ref{fig:pattern-validating-results}.
The geometrical simulation shows Kikuchi band centers and zone axes labeled by Miller indices $[uvw]$ plotted on top of the experimental EBSP.
The geometrical simulations are based on the supplementary material to the work by \textcite{britton2016tutorial}.
The list of reciprocal lattice vectors or reflectors $\{hkl\}$ used in this simulation are the four brightest families according to kinematical diffraction, $\{111\}$, $\{200\}$, $\{220\}$, and $\{311\}$.
The reflector list is created using tools available in diffsims \cite{diffsims_concept_doi}.

Since dynamically simulated patterns are used in dictionary indexing and refinement, it is instructive to visually compare the best-matching simulated pattern and the experimental EBSP.
Both a kinematical and dynamical simulation using the best-matching orientation and PC is shown in Fig. \ref{fig:pattern-validating-results}.
The dynamical simulation was projected from a master pattern generated with EMsoft.
The kinematical simulation \cite{zhu2019automated} was projected from a master pattern generated with kikuchipy.
Both simulations use the same reflector list with all allowed reflections for aluminium with a minimum interplanar spacing $d_{\mathrm{hkl}} \geq$ \SI{0.05}{\nano\meter}.
Intensities in the dynamical simulation are more realistic \cite{winkelmann2016physics}, especially in the zone axes.
Overall, the refined orientation and projection center seem valid, as a good match is seen between the experimental EBSP and the simulated EBSPs.
This type of visual validation is useful not only for the final indexing results after refinement.
It should also be used for intermediate results after the steps to optimize in Fig. \ref{fig:workflow}, such as after detector-sample geometry calibration and dictionary indexing.
If the experimental patterns do not match the simulated patterns, we should improve the current indexing step by suitable adjustment of parameters and compare again.

Another useful visual feedback is so-called inverse pole figure (IPF) maps.
In an IPF map, each map point is assigned a color given by the crystal direction $[uvw]$ parallel to a chosen sample direction.
Coloring of the IPFs for the eleven Laue point groups is supported in orix.
When satisfied, we can export indexing results to files readable by other software such as MTEX \cite{bachmann2011grain} or DREAM.3D \cite{groeber2014dream}.
Further analysis such as grain reconstruction and quantification of preferred crystal orientations, or texture, is currently unavailable in kikuchipy and orix.
It can be done in other software, such as those mentioned.
In the application examples in this work, we visually validate indexing results using both simulations and IPF maps.
Grain reconstruction and texture analysis is done with MTEX v5.9 \cite{bachmann2011grain}.


\section*{Example I: Orientation relationships in a super duplex stainless steel}
\label{sec:sdss}

Interfaces, whether they are grain or phase boundaries, play a crucial role in the control of mechanical properties of alloys.
For example, heat treatments of duplex stainless steels in the temperature range \SIrange{600}{900}{\celsius} lead to a series of phase transformations which take place mainly at ferrite/austenite ($\alpha/\gamma$) phase boundaries and inside the ferritic matrix \cite{nilsson1992super}.
The decomposition of supersaturated ferrite into phases such as secondary austenite, the intermetallic phases of sigma ($\sigma$), chi ($\chi$), $R$, and $\pi$, chromium nitrides (e.g. Cr$_2$N), and carbides (e.g. M$_{23}$C$_6$) can deteriorate the otherwise high corrosion resistance and mechanical strength of the steel.
Many of these phases precipitate with preferred crystallographic orientation relationships (ORs) to the parent phase, which can dictate the newly formed phases' growth.
Therefore, an understanding of the occurrence and prevalence of ORs is needed to maintain the desired properties of the steel.

To demonstrate characterization and visualization of ORs in terms of the misorientation with kikuchipy, we analyze an EBSD dataset from a super duplex stainless steel (SDSS) containing large amounts of undesirable secondary phases.
The steel's chemical composition is listed in Tab. \ref{tab:sdss-composition}.
The initial microstructure of approximately equal amounts of ferrite and austenite is altered during a heat treatment to promote the decomposition of ferrite.
The heat treatment consists of \SI{4}{\hour} at \SI{750}{\celsius} followed by 4 weeks at \SI{300}{\celsius}.
The dataset is acquired from a region of interest (ROI) of (224.4 $\times$ 156.8) \si{\micro\meter\squared}.
The ROI contains larger grains and a lamellar microstructure.
A step size of \SI{0.2}{\micro\meter} is chosen so as to capture many of the narrower lamellae while keeping the scanning of the relatively large area tractable.
Patterns of $120 \times 120$ px are collected at a speed of \SI{400}{\per\second} on a NORDIF UF-1100 EBSD detector in a Zeiss Ultra 55 FEG-SEM.
The accelerating voltage is \SI{20}{\kilo\volt} and the sample is positioned at a working distance of \SI{25.9}{\milli\meter} with a \SI{70}{\degree} tilt.

\begin{table}[htbp]
  \centering
  \caption{Chemical composition of the super duplex stainless steel in wt.\%.}
  \begin{tabular}{c c c c c c c c c c}
    \toprule
    Cr & Ni & Mo & Mn & Si & N & Cu & P & C & S\\
    \midrule
    24.8 & 6.6 & 3.7 & 0.85 & 0.32 & 0.26 & 0.16 & 0.023 & 0.02 & 0.0003\\
    \bottomrule
  \end{tabular}
  \label{tab:sdss-composition}
\end{table}


\subsection*{Inspection of data}

\begin{figure*}[htbp]
  \centering
  \includegraphics[width=\textwidth]{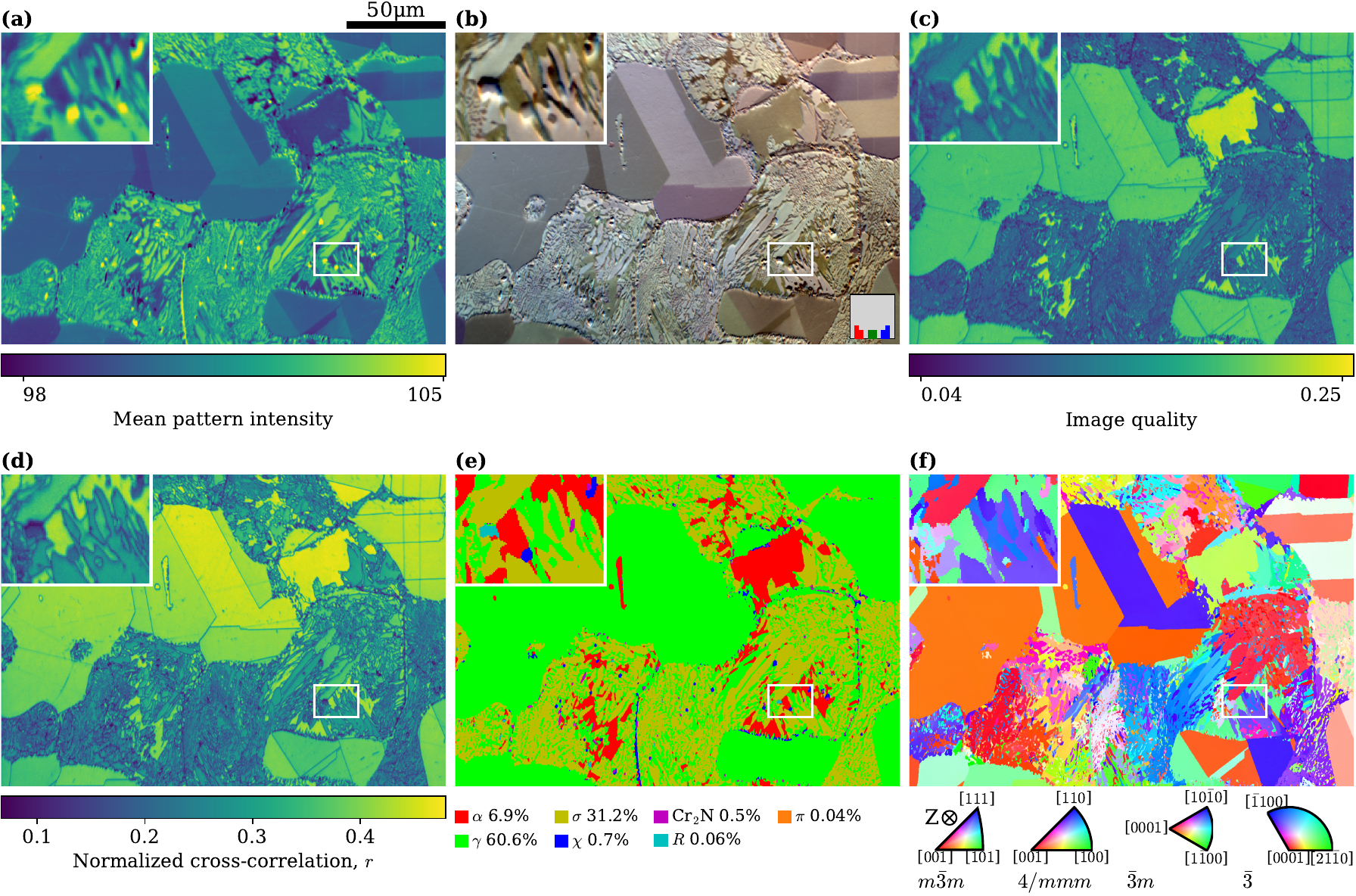}
  \caption{
  Pre-indexing (a-c) and indexing (d-f) maps from the super duplex stainless steel dataset investigated in Example I.
  (a) Mean intensity map with contrast enhanced by excluding intensities below 0.1\% and above 99.9\% in the map.
  (b) VBSE RGB image.
  The bottom right inset shows the ROIs on the detector used to obtain the red, green, and blue images that are then combined.
  (c) Image quality map.
  (d) Normalized cross-correlation score map.
  (e) Phase map.
  Phase fractions for the identified phases are shown below the map.
  (f) IPF-Z map.
  }
  \label{fig:sdss-maps}
\end{figure*}

To obtain an overview of the information in the EBSD dataset independent of indexing, we inspect three maps derived from the dataset, shown in Fig. \ref{fig:sdss-maps} (a-c).
These pre-indexing maps aid in the selection of candidate phases for indexing and the interpretation of the indexing results.
A zoomed in region is highlighted in all maps in Fig. \ref{fig:sdss-maps}, shown in insets in the upper left corners.
The first two maps are obtained by virtual backscatter electron (VBSE) imaging.
In VBSE imaging, the EBSD detector is used as a 2D BSE detector to exploit the angular distribution of the BSE yield \cite{schwarzer2011imaging,wright2015electron,nolze2017electron}.
The mean pattern intensity map in (a) represents the mean of all pixel intensities within each raw EBSP.
Its contrast, similar to that in a BSE image, is dominated by mean atomic number $Z$-contrast.
Some contrast also arises from changes in topography and crystal orientation.
The second VBSE map in (b) is a composite color image created by combining three different VBSE images in the red, green, and blue color channels.
The three VBSE images are obtained by summing the intensities within the red, green, and blue ROIs on the EBSD detector indicated in the bottom right corner of (b).
The contrast in this map depends on the chosen ROIs on the EBSD detector as well as the chosen imaging conditions \cite{nolze2017electron}.
In our case, the contrast is a combination of topography and orientation.
The final pre-indexing map in (c) is an image quality map.
It indicates the amount of high frequency content in a pattern \cite{lassen1994automated,marquardt2017quantitative} with values between 0 and 1.
It is often used as an indication of how pronounced the Kikuchi bands in a pattern are.
It is appropriate to correct the background of patterns before calculating the image quality.
Processing of the SDSS patterns comprises subtraction of a static and dynamic background as already described.

The pre-indexing maps in Fig. \ref{fig:sdss-maps} (a-c) clearly show the microstructure of larger grains and the lamellar structure.
The alternating similar intensities and colors across straight boundaries in the larger grains indicate annealing twins, common in austenite \cite{chen2001effects}.
Some lines of lower image quality in (c), such as the vertical lines in the upper right grain, do not correspond to any recognizable features in the other pre-indexing maps in (a, b).
This indicates that these are surface scratches remaining from sample preparation.
Much of the lamellar structure seems to consist of layers of higher $Z$-contrast in between layers of lower $Z$-contrast similar to that of the larger grains.
The lamellar structure is reminiscent of a eutectoid structure resulting from decomposition of supersaturated ferrite into secondary austenite and the sigma phase \cite{nilsson1993influence,lee2012isothermal}.
The sigma phase contains Cr and Mo, which gives a higher $Z$-contrast.
There are also particles of even higher $Z$-contrast dispersed in the lamellar structure.
They indicate the presence of heavier phases than ferrite, austenite, and sigma.
Good candidates are phases rich in Mo, such as the intermetallic phases chi, $R$, and $\pi$.
They are usually present in much smaller quantities than sigma.
Also, many of these particles appear round or polygonal-shaped, shapes which chi particles have been observed to precipitate in \cite{redjaimia2004morphology,lee2012isothermal}.
Finally, there are some regions inside the lamellar structure in the form of needles and bands of particles that have a lower $Z$-contrast than the grains.
These regions may be chromium nitrides Cr$_2$N, known to precipitate at about the same time as secondary austenite in steels of similar composition \cite{ramirez2003relationship}.
The chosen descriptions from literature of the crystal structures of these candidate phases are listed in Tab. \ref{tab:sdss-phases}.
The structures and atom positions given in the referenced sources are used to create dynamical simulations with EMsoft.

\begin{table*}[htbp]
  \centering
  \caption{Phase descriptions used to create dynamical simulations for indexing of the super duplex stainless steel.}
  \begin{tabular}{l l l l l l}
    \toprule
    Phase & Chemical formula & Structure & Lattice parameters [nm] & Space group & Source\\
    \midrule
    Ferrite ($\alpha$) & Fe & BCC & $a = 0.28665$ & $Im\bar{3}m$ (229) & \cite{wyckoff1963crystal}\\
    Austenite ($\gamma$) & Fe & FCC & $a = 0.3595$ & $Fm\bar{3}m$ (225) & \cite{nishihara2012isothermal}\\
    Sigma ($\sigma$) & Fe-Cr-Mo & Tetragonal & $a = 0.8799, c = 0.4566$ & $P4_2/mnm$ (136) & \cite{kiesheyer1976investigation}\\
    Chi ($\chi$) & Fe$_{36}$Cr$_{12}$Mo$_{10}$ & BCC & $a = 0.8920$ & $I\bar{4}3m$ (217) & \cite{kasper1954ordering}\\
    Chromium nitride & Cr$_2$N & Trigonal & $a = 0.48, c = 0.4472$ & $P\bar{3}1m$ (162) & \cite{lee2005crystal}\\
    $R$ & Fe-Cr-Mo & Trigonal & $a = 1.098, c = 1.948$ & $R\bar{3}m$ (148) & \cite{liu1990metastable}\\
    $\pi$ & Fe-Mo & Simple cubic & $a = 0.6315$ & $P4_132$ (213) & \cite{shoemaker1978refinement}\\
    \bottomrule
  \end{tabular}
  \label{tab:sdss-phases}
\end{table*}


\subsection*{Indexing}

For dictionary indexing, the average PC is found by indexing five so-called `calibration patterns' of $160 \times 160$ px.
The patterns are from larger grains close to the map corners and in the center.
Estimates of orientations and PCs are first found by Hough indexing and then refined using dynamical simulations.
All calibration patterns matched austenite best, with a lowest best-fit score of $r = 0.5474$.
This calibration results in an average PC of (PC$_{\mathrm{x}}$, PC$_{\mathrm{y}}$, PC$_{\mathrm{z}}$) = (0.4887, 0.3238, 0.5519).
A map of dynamic PCs is obtained by extrapolating from this average \cite{singh2017application}.
A measured detector pixel of \SI{70}{\micro\meter} for the NORDIF UF-1100 EBSD detector is used\footnote{Measured by imaging graph paper with the detector.}.
All patterns are then indexed by dictionary indexing followed by orientation refinement.
Separate runs are done for each candidate phase in Tab. \ref{tab:sdss-phases}.
The final phase map is obtained by choosing the best-fit phase and orientation in each point based on $r$.


\subsection*{Validation of indexing results}

Indexing results are presented in Fig. \ref{fig:sdss-maps} (d-f) in maps of the NCC scores $r$, phase, and IPF-Z.
The continuous variations in score values and phase and IPF-Z colors within the larger grains, lamellar structure, and particles identified in the pre-indexing maps in Fig. \ref{fig:sdss-maps} (a-c) are largely in line with expectations.
The average NCC score is $\left<r\right>$ = 0.3243, with all values in the range 0.0664--0.4793.
Points with higher scores correspond well with points of higher image quality as seen in (c).
The phase fractions of ferrite, austenite, and sigma are 6.9\%, 60.6\%, and 31.2\%, respectively.
The two latter phases constitute most of the lamellar structure.
Most of the ferrite is within the lamellar structure, with only one larger ferrite grain remaining.
This phase distribution can be described by the mentioned eutectic decomposition of ferrite.
Particles that have precipitated and grown to a large enough size to be detected by EBSD with the chosen settings constitute about 1.3\% of the area fraction with 0.7\% as chi, 0.5\% as Cr$_2$N, 0.06\% as R, and 0.04\% as $\pi$.
The locations of these particles correspond well with the locations of highest (for chi and $R$) and lowest (for Cr$_2$N) $Z$-contrast.
This is seen by comparing the maps of mean pattern intensity and phase color in Fig. \ref{fig:sdss-maps} (a) and (e), respectively.

While it is common to characterize sigma and chi with EBSD \cite{bordin2016ebsd}, it is less common for Cr$_2$N, $R$, and $\pi$.
These phases are instead studied in the transmission electron microscope due to the need for a higher spatial resolution \cite{lee2012isothermal,pettersson2015precipitation}.
This work demonstrates the ability of EBSD to characterize particles of these phases as small as \SI{0.52}{\micro\meter} in equivalent circular diameter, equivalent to five points in the dataset.


\subsection*{Analysis of orientation relationships}

To assess whether the formation of the lamellar structure or the precipitation and growth of particles follow well defined orientation relationships to the parent phase, we inspect misorientations across phase boundaries.
Phase boundaries are obtained by reconstructing grains with MTEX using a misorientation angle threshold of \SI{1}{\degree} and excluding grains smaller than five points.
ORs commonly observed in SDSS are given in Tab. \ref{tab:sdss-or}.
They are described by geometrical models expressed as parallelisms between low-index crystallographic planes and directions.
The coincident site lattice (CSL) 3 relationship is also listed.
The CSL3 relates two adjacent crystals by a \SI{60}{\degree} rotation about $\left<111\right>$.
Both \textcite{nolze2004characterization} and \textcite{krakow2017onthree} show that although ORs often can be well approximated by geometrical models, the inevitable spread in experimentally measured misorientations are better characterized in a three-dimensional misorientation space.
We therefore visualize misorientations in the axis-angle fundamental zone corresponding to each pair of phases' proper point groups \cite{morawiec1996rodrigues}.

\begin{table*}[htbp]
  \centering
  \caption{Common orientation relationships observed in super duplex stainless steel.}
  \begin{tabular}{l l l l}
    \toprule
    Phases & Orientation relationship & Definition of misorientation & Source\\
    \midrule
    $\alpha/\gamma$ or $\gamma/\chi$ & Pitsch & $(101)_{\alpha} \: || \: (010)_{\gamma}, [\bar{1}11]_{\alpha} \: || \: [101]_{\gamma}$ & \cite{nolze2004characterization}\\
    & Kurdjumov-Sachs (KS) & $(011)_{\alpha} \: || \: (111)_{\gamma}, [\bar{1}11]_{\alpha} \: || \: [\bar{1}01]_{\gamma}$ & \cite{nolze2004characterization}\\
    & Nishiyama-Wassermann (NW) & $(011)_{\alpha} \: || \: (111)_{\gamma}, [0\bar{1}1]_{\alpha} \: || \: [11\bar{2}]_{\gamma}$ & \cite{nolze2004characterization}\\
    $\alpha/\sigma$ & Chen \& Yang & $(1\bar{1}0)_{\alpha} \: || \: (\bar{1}10)_{\sigma}, [\bar{1}\bar{1}\bar{3}]_{\alpha} \: || \: [332]_{\sigma}$ & \cite{chen2001effects}\\
    $\gamma/\sigma$ & Nenno, A & $(111)_{\gamma} \: || \: (001)_{\sigma}, [\bar{1}01]_{\gamma} \: || \: [110]_{\sigma}$ & \cite{nenno1962orientation}\\
    & Chen \& Yang & $(100)_{\gamma} \: || \: (100)_{\sigma}, [011]_{\gamma} \: || \: [032]_{\sigma}$ & \cite{chen2001effects}\\
    $\alpha/\chi$ & Cube-on-cube & $(001)_{\alpha} \: || \: (001)_{\chi}, [011]_{\alpha} \: || \: [011]_{\chi}$ & \cite{redjaimia2004morphology}\\
    $\chi/\sigma$ & Redjaïmia & $(110)_{\chi} \: || \: (001)_{\sigma}, [\bar{1}11]_{\chi} \: || \: [110]_{\sigma}$ & \cite{redjaimia2004morphology}\\
    \midrule
    & CSL3 & \SI{60}{\degree} about $\left<111\right>$ &\\
    \bottomrule
  \end{tabular}
  \label{tab:sdss-or}
\end{table*}

\begin{figure*}[htbp]
  \centering
  \includegraphics[width=\textwidth]{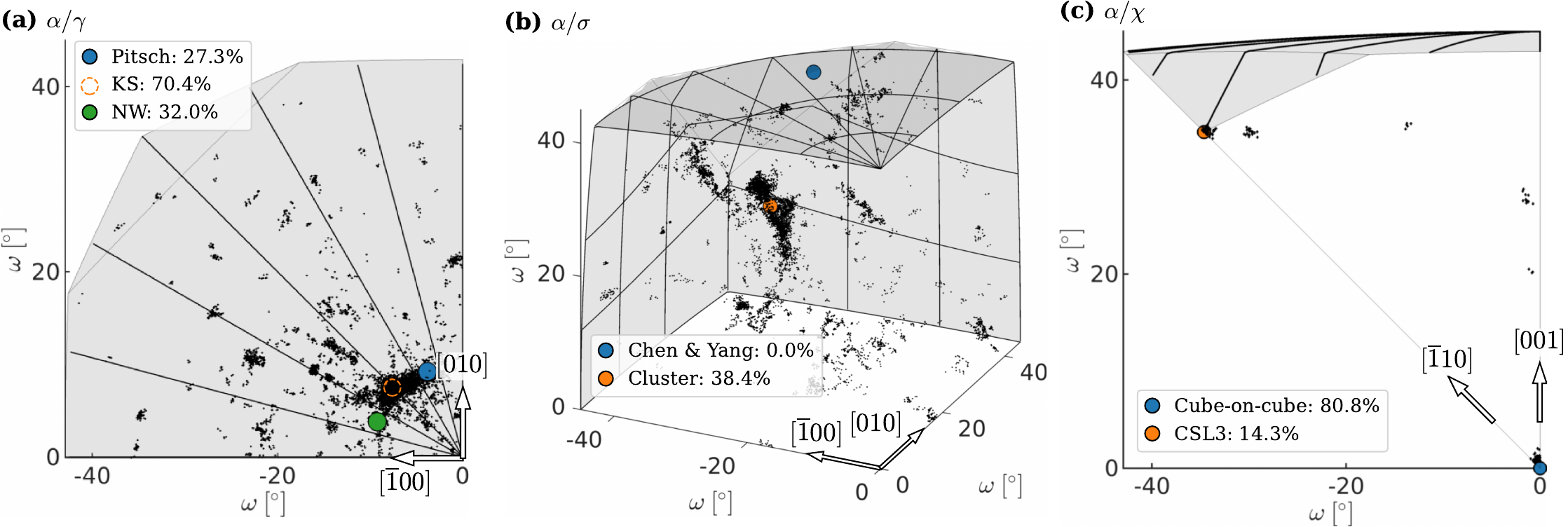}
  \caption{
  Misorientation distributions of select phase combinations in the super duplex stainless steel.
  (a) $\alpha/\gamma$ misorientations with an elongated cluster centered close to KS.
  (b) $\alpha/\sigma$ misorientations with the center of an elongated cluster highlighted.
  (c) $\alpha/\chi$ misorientations with two clusters identified as the Cube-on-cube and CSL3 misorientations.
  The fundamental zones are given by the proper point groups of the two phases in question.
  (a, c) show different projections of the 432-432 fundamental zone and (b) shows the 432-422 fundamental zone.
  The plots are produced with MTEX.
  }
  \label{fig:sdss-or-fz}
\end{figure*}

Fig. \ref{fig:sdss-or-fz} shows the distributions of misorientations across the (a) $\alpha/\gamma$, (b) $\alpha/\sigma$, and (c) $\alpha/\chi$ phase boundaries.
Relevant ORs are highlighted and labeled.
The fraction of misorientations within \SI{5}{\degree} of each OR or identified cluster is also given.
Most $\alpha/\gamma$ misorientations are clustered about the common BCC/FCC OR of Kurdjumov-Sachs (KS).
The cluster extends continuously towards the Pitsch and Nishiyama-Wassermann (NW) ORs on either side of KS.
The only systematic misorientation across austenite grain boundaries is the CSL3 relationship with 66.7\% of misorientations within \SI{5}{\degree}, confirming the appearance of annealing twins in the VBSE maps in Fig. \ref{fig:sdss-maps} (a, b).
The lack of systematic ORs across the non-CSL3 boundaries makes it difficult to identify the secondary austenite transformed from ferrite.
None of the three ORs of KS, NW, or Pitsch can therefore be ruled out as active in the transformation from ferrite to secondary austenite.
Much of the sigma phase has precipitated with a preferred OR to ferrite, with an elongated misorientation cluster visible in Fig. \ref{fig:sdss-or-fz} (b).
The cluster center is determined by taking the mean of all misorientations belonging to the biggest cluster found by density-based clustering using orix \cite{johnstone2020density}.
Clusters have a minimum of 50 points with a maximum misorientation angle of \SI{2}{\degree} between points.
39\% of misorientations are within \SI{5}{\degree} of the cluster center, which itself is \SI{0.8}{\degree} away from the low-index parallelism $(1\bar{3}\bar{1})_{\alpha} \: || \: (00\bar{1})_{\sigma}, [310]_{\alpha} \: || \: [2\bar{1}0]_{\sigma}$.
No $\sigma/\alpha$ boundaries have the OR found in a 2205 duplex stainless steel by \textcite{chen2001effects}.
They found that sigma preferentially nucleated at carbides with the FCC M$_{23}$C$_6$ crystal structure.
The carbides nucleated at $\alpha/\gamma$ phase boundaries or within ferrite grains.
No carbides are detected in the current dataset as no EBSPs match dynamical simulations of M$_{23}$C$_6$ \cite{okazaki2008effects} best.
This may explain why this particular $\alpha/\sigma$ OR is not observed.
Misorientations across $\gamma/\sigma$ boundaries are random, with only 4.2\% being within \SI{5}{\degree} of the often observed `A' OR reported by \textcite{nenno1962orientation}.
This randomness and the observed ORs between $\alpha/\gamma$ and $\alpha/\sigma$ indicate that the sigma and the austenite bordering much of the sigma in the lamellar structure have transformed from ferrite.

The chi phase is most often reported to precipitate from ferrite with a Cube-on-cube OR \cite{redjaimia2004morphology,lee2012isothermal}.
Over 200 chi particles larger than four points are detected in the EBSD dataset.
80.8\% of the chi particles' boundaries to ferrite have misorientations within \SI{5}{\degree} of the Cube-on-cube OR as shown in Fig. \ref{fig:sdss-or-fz} (c).
Interestingly, 14.3\% of misorientations have the CSL3 relationship instead.
To the best of our knowledge, this OR has not been reported before between chi and ferrite.
The significance of the CSL3-like OR for the precipitation and growth of chi particles from the ferritic matrix is not possible to determine based on this dataset alone.
Rather, it requires analysis at a higher spatial resolution and at earlier stages of precipitation.

\begin{figure*}[htbp]
  \centering
  \includegraphics[width=\textwidth]{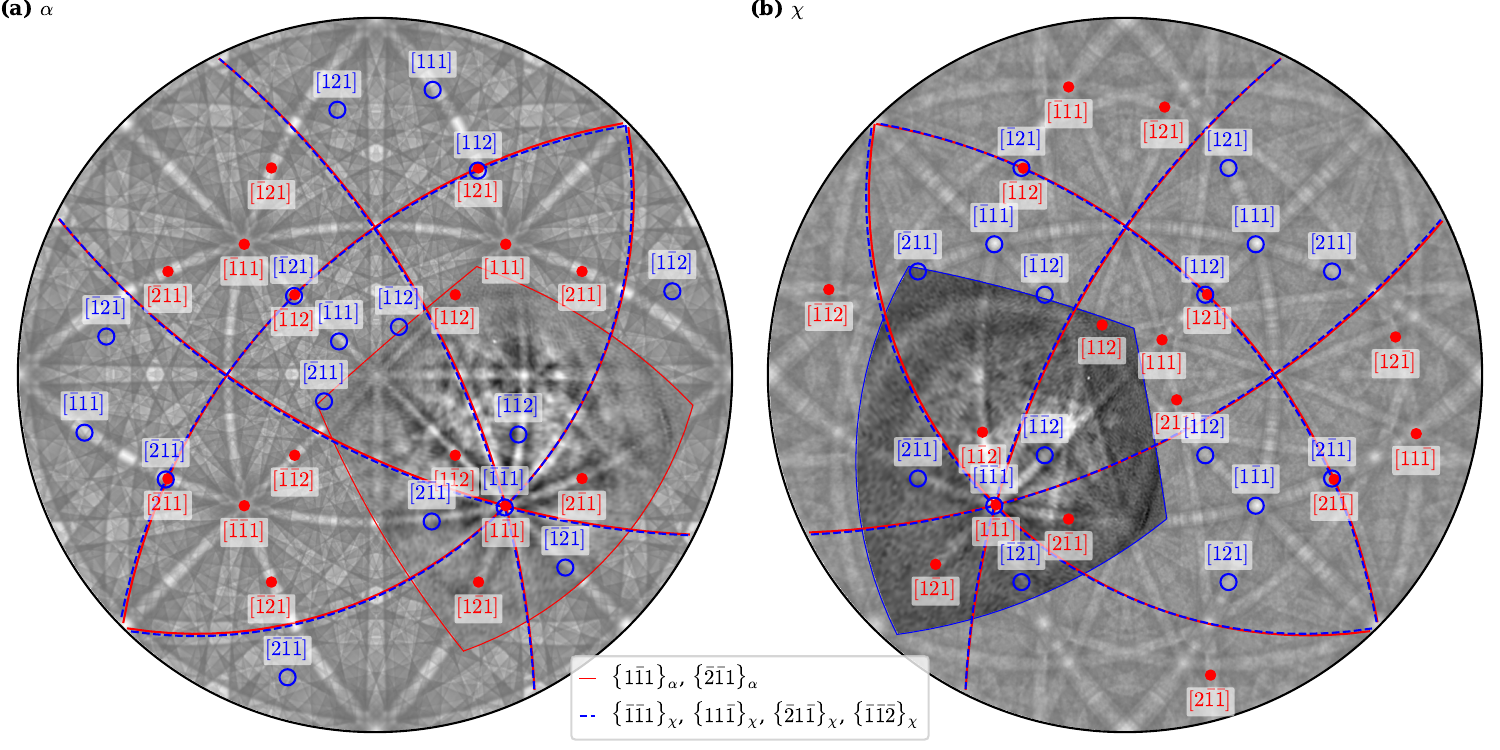}
  \caption{
  Visualization in the stereographic projection of the CSL3 OR across a phase boundary between (a) a ferrite grain and (b) a chi particle.
  Dynamical EBSD simulations of the upper hemisphere of both phases are shown slightly transparent in the background, with averaged patterns from the ferrite grain and chi particle shown as projected from the detector.
  Selected zone axes $[uvw]$ and planes $(hkl)$ particularly relevant for the CSL3 OR are shown in red full lines and circles for ferrite and blue dashed lines and open circles for chi.
  }
  \label{fig:sdss-or-fe-chi}
\end{figure*}

In the transmission electron microscope, ORs can intuitively be identified by coinciding reflections from appropriately rotated crystals of interest in selected area diffraction patterns [see e.g. work by \textcite{redjaimia2004morphology,lee2012isothermal}].
A similar representation is not immediately available using EBSPs.
Instead, ORs discerned from EBSPs are often presented in the stereographic projection as parallel planes and directions in adjacent crystals.
This representation lacks the relation back to the diffraction patterns from which the OR is determined.
To remedy this, we can back-project the EBSPs from the detector to the sphere and display them in the stereographic projection as well.
This is done in Fig. \ref{fig:sdss-or-fe-chi} for a $\alpha/\chi$ CSL3 boundary.
Parallel planes and directions of the $111$ and $211$ families of both phases are shown relative to the ferrite grain in (a) and the chi particle in (b).
The back-projected average grain and particle patterns are added on top of the upper hemispheres of the dynamically simulated master patterns used in indexing.
We see that the $(1\bar{1}1)_{\alpha}$ and $(\bar{1}\bar{1}1)_{\chi}$ planes are parallel and that all $\left<211\right>$ directions within these planes are parallel.
The poles around the $[1\bar{1}1]_{\alpha}$ zone axis visible within the ferrite EBSP show that a \SI{60}{\degree} rotation about this axis brings the crystals into coincidence, demonstrating the CSL3 OR.
In addition to being a clear visualization of an orientation relationship, Fig. \ref{fig:sdss-or-fe-chi} serves as yet another validation of the indexing results since the band positions and relative intensities in the back-projected averaged patterns fit the dynamical simulations well.

This application example demonstrates the usefulness of pre-indexing maps when gauging the information in an EBSD dataset, selecting candidate phases, and validating indexing results.
Furthermore, it shows the high fidelity and statistical significance with which we can characterize and visualize orientation relationships in a super duplex stainless steel using kikuchipy and related open-source software.


\section*{Example II: Phase differentiation in an Al--10\%Si alloy}
\label{sec:al-si}

A problem in EBSD analysis is to differentiate between phases with similar crystal structures \cite{schwartz2009electron}.
Indexing methods such as Hough indexing which rely on crystallography only---the positions and widths of Kikuchi bands, for phase differentiation---disregard the often-crucial information encoded in relative intensities between bands.
These relative intensities arise mostly from the chemical composition of a phase, meaning the atomic positions in the lattice and elemental occupancies in these positions.
An industrially important and fundamental material group in which phase differentiation using EBSD is challenging is Al--Si alloys.
Aluminium has an FCC lattice with the space group symmetry $Fm\bar{3}m$ (225) and a lattice parameter of $a =$ \SI{0.405}{\nano\meter} \cite{wyckoff1963crystal}.
Silicon has a diamond cubic lattice with the space group symmetry $Fd\bar{3}m$ (227) and a lattice parameter of $a =$ \SI{0.543}{\nano\meter} \cite{wyckoff1963crystal}.


\subsection*{Similarities in simulated EBSPs from aluminium and silicon}

\begin{figure*}[htbp]
  \centering
  \includegraphics[width=\textwidth]{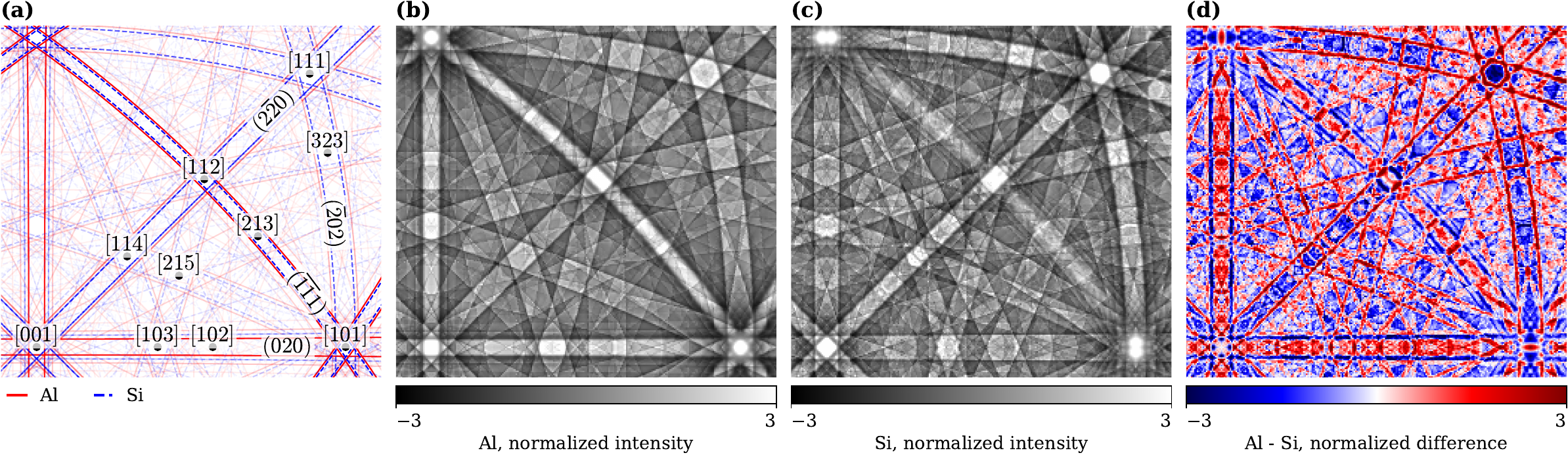}
  \caption{
  Differences in simulated EBSPs from aluminium and silicon at \SI{20}{\kilo\volt}.
  The $m\bar{3}m$ fundamental sector is shown.
  (a) Geometrical simulation of aluminium (full red line) and silicon (dashed blue line).
  Bright low-index zone axes $[uvw]$ and bands $(hkl)$ for aluminium are highlighted.
  (b, c) Normalized dynamical simulations of aluminium and silicon.
  (d) Difference pattern between (b) and (c).
  Red (blue) corresponds to stronger aluminium (silicon) scattering.
  }
  \label{fig:al-si-similarity}
\end{figure*}

The similarity in the crystallography of aluminium and silicon becomes apparent when considering geometrical Kikuchi pattern simulations.
Fig. \ref{fig:al-si-similarity} (a) shows the fundamental sector of both phases' Laue class $m\bar{3}m$ in the stereographic projection.
The 15 brightest Bragg reflections from each phase are shown as bands delineated by Friedel pairs $(hkl)$ and $(\bar{h}\bar{k}\bar{l})$ spaced $2\theta$ apart (given by an accelerating voltage of \SI{20}{\kilo\volt}).
Their color is scaled by the structure factor $|F_{\mathrm{hkl}}|$ using kinematical atomic scattering parameters \cite{prince2004international}.
Some bright low-index zone axes $[uvw]$ and bands $(hkl)$ for aluminium are labeled.
It is clear why the band centers of the strongest bands alone may be insufficient to differentiate between aluminium and silicon, as they are in the same positions.
The band widths, on the other hand, differ due to the relative difference in the lattice parameter of 25\%.
Even so, \textcite{liu2015twin} obtained an unsatisfactory differentiation in a eutectic Al--12.7\%Si alloy using a commercial Hough indexing software that accounts for band widths.
A common solution to this kind of problem is to simultaneously collect an EBSP and an energy dispersive X-ray spectrum (EDS) from each mapping point.
The spectrum can be used to reduce the list of candidate phases for the EBSP.
There are however several challenges in this correlative approach due to limitations specific to each technique as noted by \textcite{nowell2004phase}, e.g. the lower ultimate spatial resolution of EDS ($\sim$\SIrange{1}{2}{\micro\meter}) compared to EBSD ($\sim$\SI{0.01}{\micro\meter}).

\textcite{ram2018phase} show that phases with similar crystal structures can be reliably differentiated with EBSD alone by dictionary indexing.
As opposed to Hough indexing, dictionary indexing with dynamical simulations use the complete intensity distribution in a pattern to determine the best-fit phase, orientation, and other features of interest.
Parts of the dynamical simulations of aluminium and silicon, including the fundamental sector, are shown in Fig. \ref{fig:al-si-similarity} (b, c), respectively.
The simulations take into account reflections with a minimum interplanar spacing of $d_{\mathrm{hkl}} \geq$ \SI{0.05}{\nano\meter}.
To allow direct comparison of relative intensities, we have normalized both simulations to a mean of 0 and a standard deviation of 1 before visualization.
We recognize many of the strongest zone axes and bands from the kinematical simulations in (a).
Clear differences are visible in band widths and intensities between the two phases.
These differences result in a similarity score between the two simulation parts of $r = 0.3256$.
How the relative intensities of the phases' simulations differ is easiest to gauge in the difference pattern in (d).
The difference pattern is obtained by subtracting the normalized silicon from the normalized aluminium simulation.
Great differences are seen in the $\{200\}$ ($\{400\}$ in silicon), $\{220\}$, and $\{111\}$ bands delineating the fundamental sector.
Aluminium has brighter and narrower $\{200\}$ ($\{400\}$) and $\{111\}$ bands than silicon, seen as red delineated by bright blue, while the opposite is true for the $\{220\}$ bands.


\subsection*{Indexing}

Whether dictionary indexing followed by refinement using dynamical simulations can distinguish between aluminium and silicon based on EBSPs alone is tested on patterns from a hypoeutectic Al--10\%Si alloy.
The alloy is modified with 0.018 wt.\% Sr and 0.04 wt.\% Ti.
The sample has a typical microstructure for modified hypoeutectic Al--Si alloys with silicon particles as clustered, rounded fibers of sub-micron size in-between primary aluminium grains \cite{makhlouf2001aluminum}.
To collect patterns from the particles, a step size of \SI{0.02}{\micro\meter} is used.
Patterns of $156 \times 88$ px are acquired at \SI{740}{\per\second} in a rectangular ROI covering $301 \times 251$ points in an area of 6.02 $\times$ 5.02 \si{\micro\meter\squared}.
The patterns are collected on an Oxford Instruments Symmetry S2 EBSD detector in a Zeiss FEG-SEM operated at \SI{20}{\kilo\volt}.
The working distance is \SI{10.9}{\milli\meter} and the sample is tilted to \SI{70}{\degree} while the detector is tilted \SI{5}{\degree} away from the sample.

Using an average PC for both dictionary indexing and refinement is valid as the ROI is small.
The average PC is found by indexing ten manually selected aluminium patterns from the dataset.
Estimates of orientations and PCs for these patterns are first found by Hough indexing and then refined simultaneously using dynamical simulations.
A few of the patterns have a relatively low similarity of $r < 0.4$ to the best-matching simulations, and so are excluded when calculating the average PC.
The average PC is (PC$_{\mathrm{x}}$, PC$_{\mathrm{y}}$, PC$_{\mathrm{z}}$) = (0.4986, 0.0457, 1.2177).
The final phase map is obtained by choosing the best-fit phase and orientation in each point based on $r$.


\subsection*{Validation of indexing results}

\begin{figure*}[htbp]
  \centering
  \includegraphics[width=\textwidth]{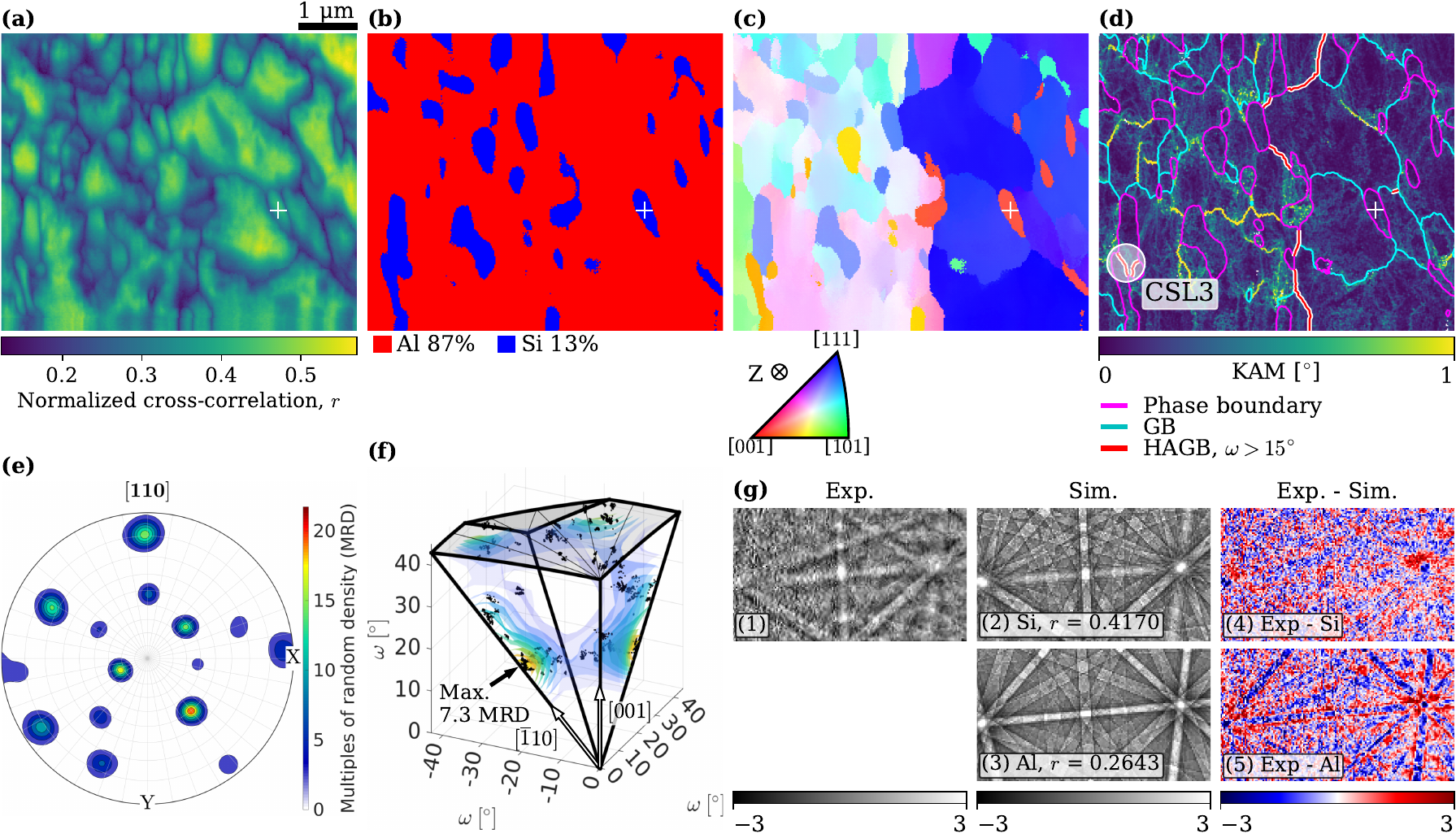}
  \caption{
  Indexing results from a modified Al--10\%Si alloy.
  (a) Normalized cross-correlation score map.
  (b) Phase map.
  (c) IPF-Z.
  (d) KAM map with phase and grain boundaries highlighted according to the inset legend.
  (e) Pole density function of $\left<110\right>$ directions in the Si particles.
  (f) Al-Si misorientations in axis-angle space, with the maximum highlighted.
  (g) Best-fit Al and Si simulations to an experimental pattern highlighted with a cross in (a-d).
  (d, e, f) are made with MTEX.
  }
  \label{fig:al10si}
\end{figure*}

The average similarity score after refinement is $\left<r\right> = 0.3402$.
All scores are in the range \SIrange{0.12}{0.57}{}.
Scores are in general lower at grain, particle, or phase boundaries and higher within particles or grains, as seen in Fig. \ref{fig:al10si}(a).
Careful scrutiny of the scores is important before analyzing the indexing results.
The distribution of scores (not shown) is continuous, and there is no obvious threshold score to identify incorrect indexing solutions by.
In challenging cases, information independent of indexing should be used to identify incorrect solutions.
Such information can be maps with topographic contrast to identify points without diffracting signal.
Pre-indexing maps such as the image quality map \cite{lassen1994automated} or a virtual backscatter electron image \cite{wright2015electron}, used in the inspection of the SDSS dataset and shown in Fig. \ref{fig:sdss-maps}, can be useful.
The pre-indexing maps for this particular dataset do not produce any meaningful contrast.

Instead of considering these maps, we can further validate our results by comparing the best-matching simulated silicon and aluminium patterns to an experimental EBSP.
The experimental EBSP, highlighted in Fig. \ref{fig:al10si} (a) with a cross, is indexed as silicon.
The experimental EBSP and the best-matching simulated silicon and aluminium patterns are shown in (g-1), (g-2), and (g-3), respectively.
The best-fit score $r$ for silicon is about 37\% greater than for aluminium.
The lower match to aluminium is mostly because of the mismatch between experimental and simulated band intensities and widths for aluminium, seen as negative values within bands in the difference pattern in (g-5).
The band intensities and widths in the silicon simulation, on the other hand, correspond well to the experimental pattern.
This is seen as a lack of strongly correlated positive or negative values in the difference pattern in (g-4).

\subsection*{Orientation analysis of silicon fibers}

The microstructure of silicon fibers is recognizable in the phase map in Fig. \ref{fig:al10si} (b).
13\% of patterns match silicon best.
The slightly higher silicon fraction than the 10 wt.\% in the alloy seems reasonable since silicon particles are inhomogeneously distributed in between the primary aluminium dendrites.
The solubility of each element in the other is also low.
The IPF-Z map in (c) shows that the ROI covers roughly part of two primary aluminium dendrites containing several subgrains.
The map also shows that the silicon particles are textured.
The first observation is confirmed after reconstructing grain boundaries (GB) and phase boundaries with MTEX using a misorientation angle threshold of \SI{1}{\degree}.
These boundaries are overlayed on the kernel average misorientation (KAM) map in (d).
The only high-angle grain boundaries (HAGBs), defined as having a misorientation angle $\omega >$ \SI{15}{\degree}, separate the two primary aluminium grains and three silicon particles presumably part of the same fiber [highlighted in a circle in (d)].
The HAGBs in the silicon fiber are the only Si--Si boundaries in the map.
The misorientations across these boundaries are within \SI{3}{\degree} of a CSL3 twin misorientation.
These twins are important for growth of silicon flakes and fibers \cite{day1968microstructure} and for fiber branching \cite{shamsuzzoha1986crystal} during solidification of Al--Si alloys.
The formation of the three silicon particles making up the fiber can be described by low-angle fiber branching.

Nearly all silicon particles share a $\left<110\right>$ direction.
This is clear from the $>20$ multiples of random density (MRD) maximum in the pole density function plotted in the stereographic projection in Fig. \ref{fig:al10si} (e).
The pole density function is calculated by kernel density estimation with a kernel half-width of \SI{5}{\degree} \cite{hielscher2008novel}.
This $\left<110\right>$ texture was observed in both unmodified and Sr-modified Al--12.7\%Si alloys by \textcite{liu2015twin}.

Fig. \ref{fig:al10si} (f) shows the distribution of misorientation between aluminium grains and silicon particles in the 432--432 fundamental zone in axis-angle space.
All misorientations and the misorientation distribution function are shown.
The misorientation distribution function is calculated by kernel density estimation with a kernel half-width of \SI{5}{\degree}.
No clear orientation relationship is visible.
The maximum in the distribution is about 7.3 MRD and corresponds to a \SI{28}{\degree}$\left<110\right>$ misorientation.
However, only about 15\% of misorientations are within \SI{5}{\degree} of this maximum.
The lack of an Al/Si orientation relationship in Sr-modified eutectic Al--Si alloys is a common observation \cite{shamsuzzoha1986crystal,liu2015twin}.

This application example shows that EBSD alone can differentiate between aluminium and silicon.
Furthermore, it demonstrates that comparing simulations is a powerful tool when evaluating challenges in phase differentiation.
Lastly, it is stressed that comparing experimental to best-matching simulated EBSPs should be done to validate indexing results.
kikuchipy provides the necessary tools to reach these insights.


\section*{Example III: Phase differentiation of constituent particles in an Al-Mn alloy}
\label{sec:al-mn}

The final example concerns indexing of second-phase particles in an Al--Mn alloy from noisy EBSPs.
The particles of interest are micron-sized constituent particles formed during direct-chill casting of an alloy with a composition of Al--0.53Fe--0.39Mn--0.152Si wt.\%.
The main particle phases in Al--Mn alloys are Al$_6$Mn and $\alpha$--AlMnSi.
Iron may substitute manganese in both phases.
This is ignored here, however, as the effect of substitution on dynamical EBSD simulations is negligible.
Al$_6$Mn particles form during alloy casting.
They have the orthorhombic space group symmetry $Cmcm$ (63) and lattice parameters of $(a, b, c)$ = (0.75551, 0.64994, 0.88724) \si{\nano\meter}, according to \textcite{kontio1981new}.
$\alpha$--AlMnSi particles form during the subsequent heat treatment of the alloy.
They can take on a primitive cubic or a BCC structure depending on the Mn/Fe ratio \cite{donnadieu1994manganese}.
The $\alpha$--AlMnSi particles are assumed to have the primitive cubic structure in the investigated alloy, with space group symmetry $Pm\bar{3}$ (200) and a lattice parameter of $a$ = \SI{1.268}{\nano\meter} \cite{cooper1966crystal}.

The characterization challenge in this example comes from noise in the EBSPs in the investigated dataset.
The noise originates mainly from two sources.
Firstly, the Al--Mn alloy is heavily cold-rolled to 95\% reduction (a true strain of $\varepsilon$ = 3) and annealed to just before the onset of recrystallization \cite{aanes2022correlated}.
The microstructure consists of fine aluminium subgrains of average size \SI{1.55(5)}{\micro\meter} with bands of constituent particles in the rolling direction (RD).
The subgrains, especially in the immediate vicinity of the larger constituent particles, accommodate stored energy from cold-rolling as localized plastic deformation.
The stored energy results in a reduced sharpness of the Kikuchi bands in the acquired EBSPs.
Secondly, the initial motivation for acquiring the EBSD dataset was to analyze the aluminium subgrains \cite{aanes2022correlated,aanes2023orientation}.
The EBSD acquisition parameters were thus chosen to be sufficient for indexing of aluminium EBSPs.
However, this resulted in additional on-camera-related noise.


\subsection*{Indexing}

The dataset was acquired on a NORDIF UF-1100 EBSD detector in a Zeiss Ultra 55 FEG-SEM.
The accelerating voltage was \SI{17}{\kilo\volt}, the step size \SI{0.1}{\micro\meter}, and the sample was positioned at a working distance of \SI{24.4}{\milli\meter} with a \SI{70}{\degree} tilt.
The on-camera-related noise is assumed to come from the low detector pixel resolution of $96 \times 96$ px, relatively fast frame rate of \SI{70}{\per\second}, and 7 dB camera gain.
The ROI covers a nominal area of $92 \times 92$ \si{\micro\meter\squared}.
Processing of patterns include subtraction of the static and dynamic backgrounds.
To increase the SNR, each pattern is averaged to their eight nearest neighbors using a $3 \times 3$ Gaussian kernel of $\sigma$ = 1 centered on the pattern to average.

The average PC is found by indexing five calibration patterns of $240 \times 240$ px.
The calibration patterns are from aluminium subgrains close to the map corners and in the center.
Estimates of orientations and PCs are first found by Hough indexing and then refined simultaneously using dynamical simulations.
The lowest best-fit score after refinement is $r = 0.3833$.
This calibration results in an average PC of (PC$_{\mathrm{x}}$, PC$_{\mathrm{y}}$, PC$_{\mathrm{z}}$) = (0.4968, 0.1882, 0.5532).
A map of dynamic PCs is obtained by extrapolating from this average \cite{singh2017application} using a detector pixel size of \SI{70}{\micro\meter}.
All patterns are then indexed by dictionary indexing followed by orientation refinement.
Separate runs are done for aluminium, Al$_6$Mn, and $\alpha$--AlMnSi.
Particle boundaries are obtained after indexing via grain reconstruction with MTEX using a misorientation angle threshold of \SI{1}{\degree}.
To reduce potential errors from mis-indexing of the noisy EBSPs, particles smaller than five points are excluded from the analysis.


\subsection*{Validation of indexing results}

Our goal here is to analyze orientations from correctly indexed particles while at the same time excluding incorrectly indexed particles from the analysis.
We can use a `ground truth' map of particle locations to filter the particles, explained in detail by \textcite{aanes2022correlated}.
The map is obtained by thresholding a BSE image.
The BSE image shows an area fraction of constituent particles of 0.6\%, a number density of \SI{7(3)e3}{\per\milli\meter\squared}, and an average size of \SI{1.36(9)}{\micro\meter}.
The particle map is inserted into the EBSD dataset using image registration\footnote{The BSE image is warped to fit the EBSD map to avoid having to interpolate pattern intensities. Note that this transformation is in the opposite direction compared to how it was done by \textcite{aanes2022correlated}.} to obtain a multimodal dataset.
Assuming that both the particle detection and the image registration are good, the multimodal dataset shows in which EBSPs we should expect to find diffracted intensities from particles.
For consistency with the EBSD phase map, detected particles smaller than five points in the inserted particle map are excluded from the analysis.
A one-to-one comparison between the EBSD phase map and the particle map is now possible.
We can thus assess the ability of pattern matching to correctly identify constituent particles in the Al–Mn alloy from noisy EBSPs.

\begin{figure*}[htbp]
  \centering
  \includegraphics[width=\textwidth]{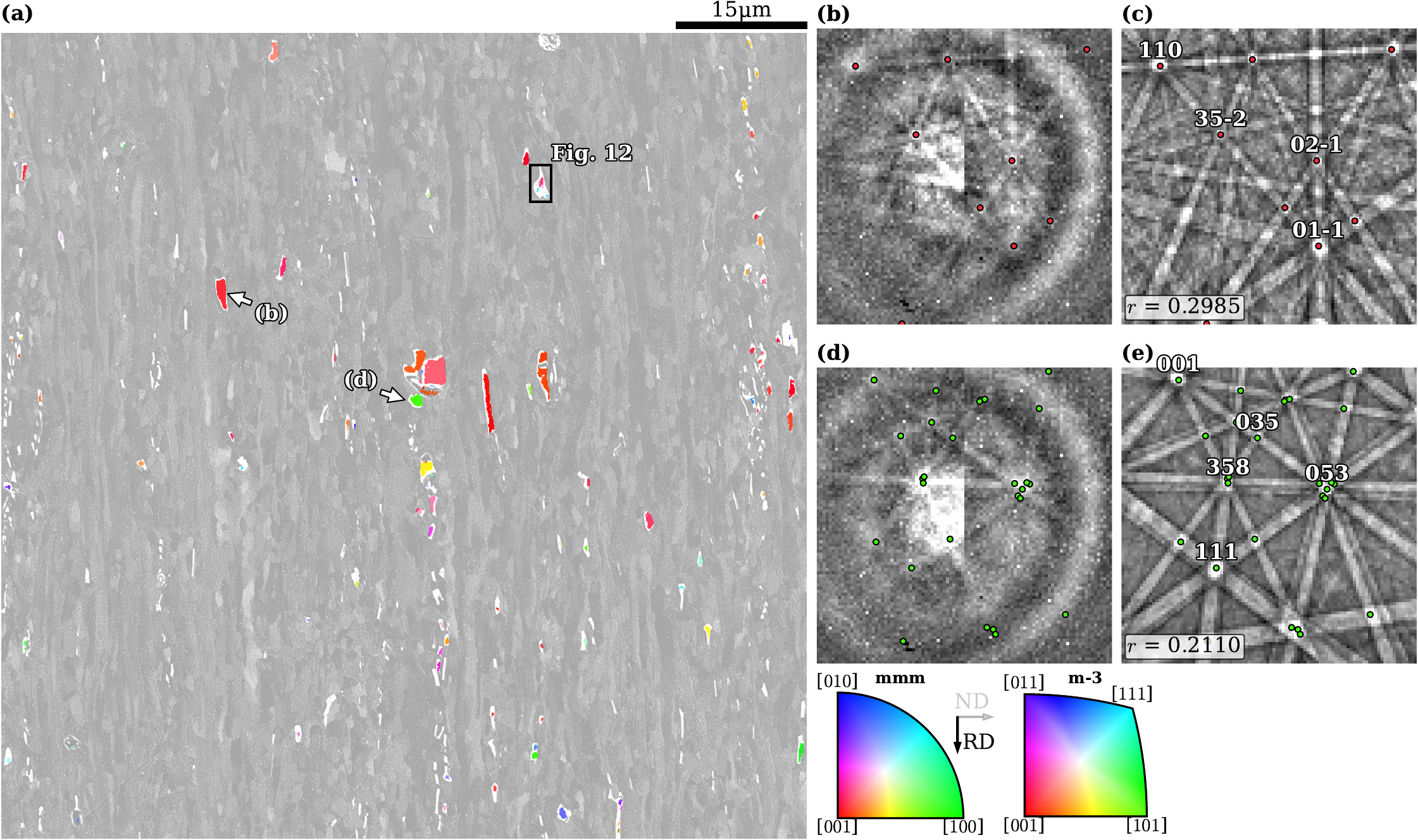}
  \caption{
  Indexed constituent particles in the Al--Mn alloy.
  (a) IPF-RD map, with colors given by the corresponding IPFs.
  The map is overlayed on a BSE image warped to fit the EBSD map.
  (b) Averaged pattern from the points within a highlighted particle in (a).
  (c) Best-fit simulation to (b), which is of Al$_6$Mn.
  (d) Averaged pattern from the points within a highlighted particle in (a).
  (e) Best-fit simulation to (d), which is an of $\alpha$--AlMnSi.
  Some zone axes $uvw$ are added as visual aid when relating features in the averaged and simulated patterns.
  }
  \label{fig:almn-particles}
\end{figure*}

Fig. \ref{fig:almn-particles} (a) shows the IPF-RD map after excluding aluminium and small particles.
The IPF-RD map is overlayed on the BSE image used in particle detection.
63 Al$_6$Mn particles and 48 $\alpha$--AlMnSi particles are indexed.
They constitute 0.43\% and 0.12\% of the map area, respectively.
The average NCC score is $\left<r\right>$ = 0.1331, with values between 0.0736--0.2261.
20\% of particle points in the particle map are indexed as particles.
Most of the remaining 80\% of particle points belong to smaller and narrower particles, as seen in Fig. \ref{fig:almn-particles} (a).
Presumably, these particles gave off too weak a signal compared to the surrounding aluminium.
The fraction of missed particle points may perhaps be reduced by using a smaller EBSD step size and a longer exposure time.
Only 6\% of points indexed as particles do not correspond to a particle in the particle map.
Visual inspection shows that most of these points border detected particles, indicating that they would match the detected particles with an improved image registration.
Still, this low fraction indicates that the image registration \cite{aanes2022correlated} provides a plausible overlap between the BSE particle map and the EBSD map.

To validate the indexing results, averaged patterns from an Al$_6$Mn and an $\alpha$--AlMnSi particle, highlighted in Fig. \ref{fig:almn-particles} (a), are shown in (b) and (d), respectively.
The sharp intensity change across a vertical line in the middle of the patterns arises from different illumination of the two CCD detector chips.
The outer ring of raised intensity is an artifact from the static background subtraction.
The simulated patterns from the average orientation within the Al$_6$Mn and $\alpha$--AlMnSi particle and the corresponding NCC scores are shown in (c) and (e), respectively.
Some prominent zone axes $uvw$ are added as visual aid when relating features in the averaged and simulated patterns.
The correspondence indicates a correct indexing.


\subsection*{Analysis of particle orientations}

\begin{figure}[htbp]
  \centering
  \includegraphics[width=\columnwidth]{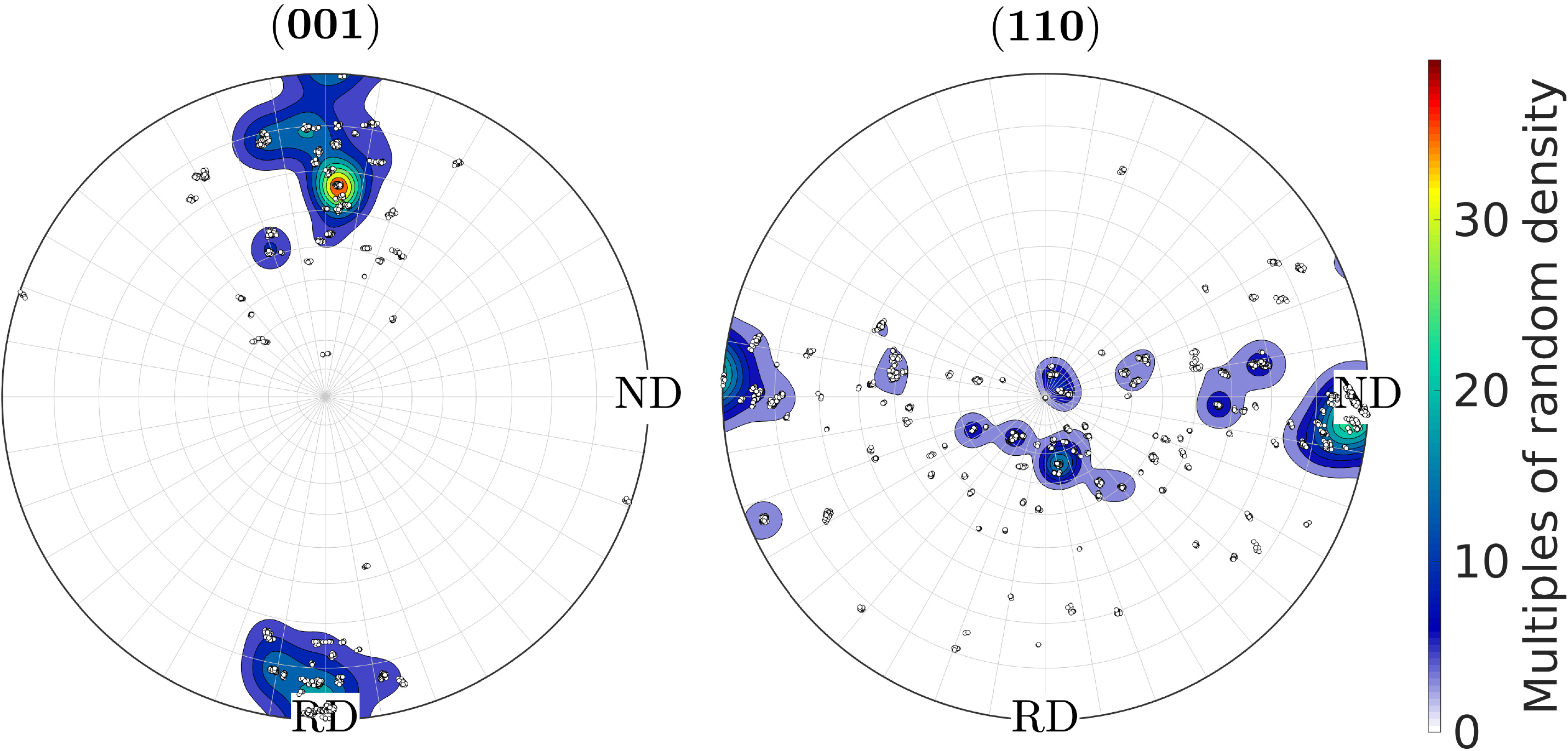}
  \caption{
  Pole density functions in the stereographic projection showing preferred directions for the perpendicular (left) $\{001\}$ and (right) $\{110\}$ plane normals in the Al$_6$Mn particles in the Al-Mn alloy.
  The plots are made with MTEX.
  }
  \label{fig:al6mn-texture}
\end{figure}

The similar orientation colors for many of the particles in Fig. \ref{fig:almn-particles} (a) suggest a preferred orientation.
The RD of the material is parallel to the cylindrical axis of the cast billet.
Most of the Al$_6$Mn particles have their $c$-axis close to parallel to this axis, while there is no equally preferred sample direction perpendicular to this axis.
This can be seen from the pole density functions of the perpendicular $\{001\}$ and $\{110\}$ crystal directions in Fig. \ref{fig:al6mn-texture}.
The maximum of the $\{001\}$ distribution is misoriented from the rolling direction by about \SI{20}{\degree}, which might be explained by the breaking up and rotation of particles during cold-rolling.
The $\alpha$--AlMnSi particles, on the other hand, have no preferred orientation.

\begin{figure*}[htbp]
  \centering
  \includegraphics[width=\textwidth]{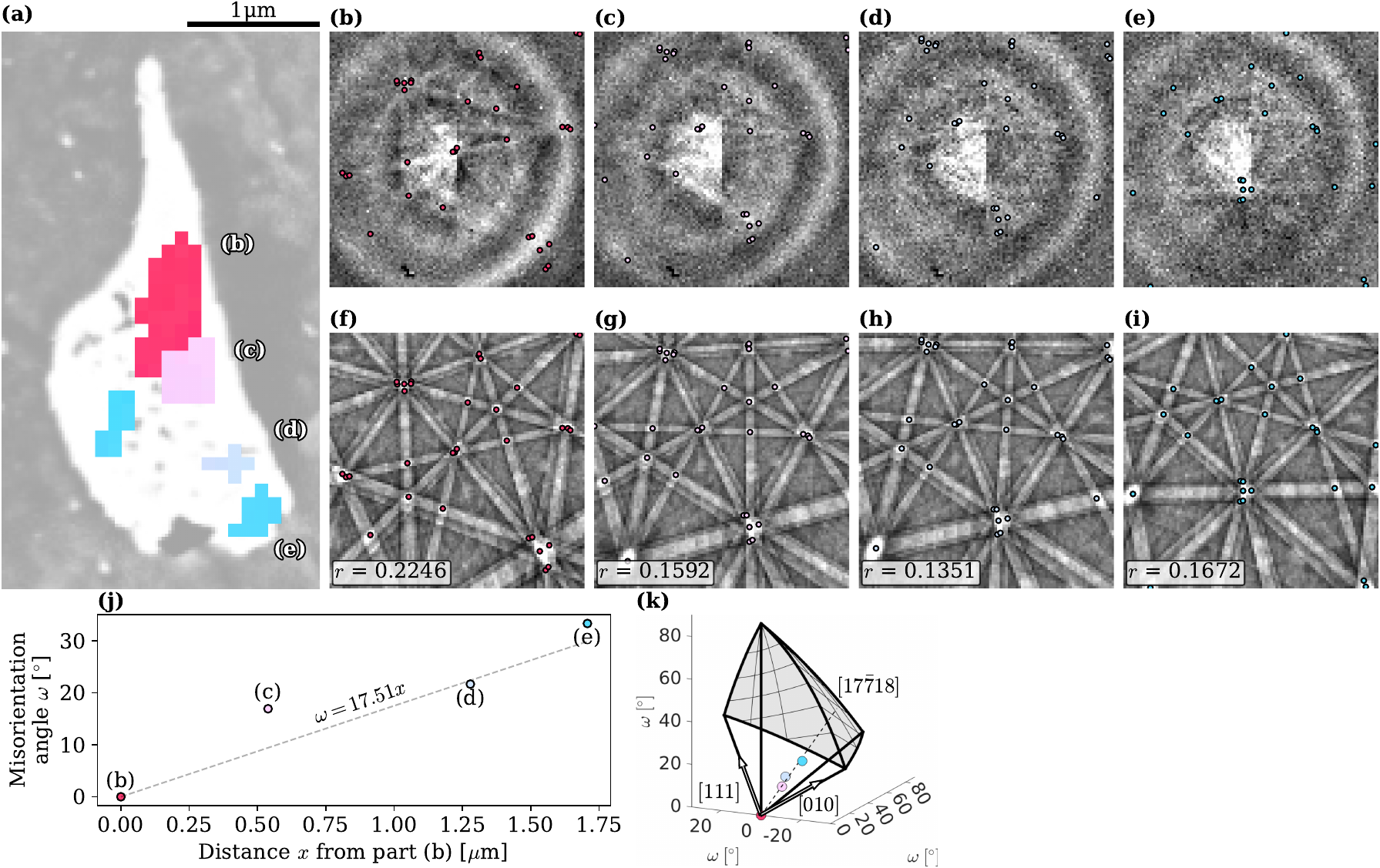}
  \caption{
  Large lattice rotation within an $\alpha$--AlMnSi particle in the Al-Mn alloy.
  (a) IPF-RD map for five detected parts of the same particle, obtained using the $m\bar{3}$ IPF key in Fig. \ref{fig:almn-particles}.
  The map is overlayed on a (transparent) BSE image warped to fit the EBSD map.
  (b-e) Averaged patterns from the parts highlighted in (a).
  (f-i) Best-fit simulations to (b-e), which are all of $\alpha$--AlMnSi.
  Some zone axes are highlighted as visual aid when relating features in the averaged and simulated patterns.
  (j) Misorientation angle $\omega$ as a function of distance $x$ from the top part (b) of the particle to the bottom part (e).
  (k) Misorientations with respect to the top part (b), plotted in axis-angle space using MTEX.
  The mean axis of rotation $[17, \bar{7}, 18]$ for parts (c-e) is highlighted.
  }
  \label{fig:almn-rotated-particle}
\end{figure*}

One particular $\alpha$--AlMnSi particle, highlighted in the square in Fig. \ref{fig:almn-particles} (a), is found to contain a large lattice rotation.
Fig. \ref{fig:almn-rotated-particle} (a) shows a magnified view of this particle.
Five parts of the particle are correctly indexed.
The average patterns from four of these parts, going from top to bottom of the particle, are shown in (b-e).
The best matching simulated patterns and corresponding NCC scores shown in (f-i).
The simulated patterns further exemplifies the lattice rotation.
To investigate the lattice rotation, we analyze the misorientations of (c-e) relative to the top part (b).
The change in the misorientation angle $\omega$ as a function of the distance $x$ from the centroid of (b) is shown in Fig. \ref{fig:almn-rotated-particle} (j).
It seems to follow a linear relationship as indicated by the trend line $\omega = 17.51x$.
The misorientations are plotted in the axis-angle fundamental zone of the proper point groups 23--23 in (k).
The axes of rotation are seen to deviate only slightly from $[17, \bar{7}, 18]$.
This systematic lattice rotation is most likely imposed on the particle from the heavy deformation.

This final application example demonstrates the ability to identify second-phase particles from noisy EBSPs in a heavily deformed Al--Mn alloy.
The flexibility with which the indexing results can be validated and analyzed is shown by a comparison to independently obtained data.


\section*{Current limitations and future improvements}

There are several improvements and advancements kikuchipy can leverage in future versions \cite{zemanek1983algorithmic}.
One particular drawback in the workflow (see Fig. \ref{fig:workflow}) is that dictionary indexing and refinement has to be done once per phase.
This requires extra setup by the user and reduces the ease-of-use.
This drawback is also present for geometrical simulations and their visualization [as shown in Fig. \ref{fig:pattern-validating-results} (a)].
The approach taken in PyEBSDIndex \cite{rowenhorst2024fast} with all phase information encoded in a single `indexer' object should be explored.

Another drawback in the workflow is the need to import dynamical EBSD simulations created by other software, such as EMsoft \cite{jackson2019dictionary}.
Having the ability to combine a crystal database with on-the-fly dynamical EBSD simulations within a Python workflow will open up several new applications within open-source EBSP analysis.
Examples include searching for a best match for a pattern with an a priori unknown phase or estimating lattice distortions by finding a best match to simulations with varying lattice parameters \cite{cios2023mapping}.

Although we with kikuchipy focus on flexibility rather than a high indexing speed, increasing the indexing speed would greatly improve the ease-of-use.
Currently, the graphical processing unit (GPU) is only used indirectly via Hough indexing with PyEBSDIndex \cite{rowenhorst2024fast}, with remaining processing being done on the central processing unit (CPU).
Pattern processing and dictionary indexing lends itself well to processing on the GPU, and should be explored.

Currently, no GUI is provided, which may make some EBSD users hesitant to try kikuchipy.
However, the popularity of HyperSpy and the MTEX texture toolbox \cite{bachmann2011grain}, with scripting as their primary interface\footnote{Although HyperSpy's core functionality is available in the HyperSpyUI \cite{pena2017electron}.}, shows that a GUI is not required to appeal to a broad user base within the materials science and geology community.


\section*{Concluding remarks}

We have presented kikuchipy, an open-source toolbox for analysis of electron backscatter diffraction patterns, written in Python.
The software's capabilities and the flexible and iterable indexing workflow was described in detail.
The workflow includes loading and inspection of data, suitable processing of pattern intensities, determination of candidate phases for indexing, calibration of the detector-sample geometry, indexing, and final validation of indexing results.
We applied this workflow in three application examples: analysis of orientation relationships in a super duplex stainless steel, phase differentiation of aluminium and silicon in a cast modified Al--Si alloy, and phase differentiation of particles in an Al--Mn alloy as Al$_6$Mn or $\alpha$--AlMnSi.

The kikuchipy source code is hosted on GitHub: \href{https://github.com/pyxem/kikuchipy}{https://github.com/pyxem/kikuchipy}.
The software documentation accessed at \href{https://kikuchipy.org}{https://kikuchipy.org} includes a complete description of all available functionality, and, more importantly, detailed tutorials and examples for users to adapt to their own analysis.
Lastly, the raw EBSD datasets and the indexing and analysis workflows used to produce the results in the application examples are available from Zenodo at \href{https://doi.org/10.5281/zenodo.20290635}{https://doi.org/10.5281/zenodo.20290635} and GitHub at \href{https://github.com/hakonanes/kikuchipy-paper-examples}{https://github.com/hakonanes/kikuchipy-paper-examples}, respectively.








\section*{Acknowledgments}

H.W.Å. acknowledges NTNU for support through the NTNU Aluminium Product Innovation Center (NAPIC).
T.B. acknowledges the support by the Research Council of Norway to SFI Manufacturing (project no. 237900).
K.M. acknowledges the support by the Research Council of Norway to SFI PhysMet (project no. 309584)
Drs. Yingda Yu, Sergey Khromov, and Morten P. Raanes are thanked for maintaining the instruments at EMlab at NTNU.
Part of the EBSD indexing was done on the NTNU Idun computing cluster \cite{sjalander2019epic}.
Dr. Odd M. Akselsen, then at SINTEF, and Vetle R. Østerhus are acknowledged for providing the SDSS sample and acquiring the EBSD dataset, respectively.
Nora Dahle at SINTEF and Ingvild Runningen, then at NTNU, are acknowledged for providing the Al--10\%Si alloy sample and preparing it for EBSD, respectively.
Dr. Grzegorz Cios at AGH University of Kraków is thanked for discussions leading to the identification of particles of the $\pi$ phase in the SDSS.
Hydro Aluminium is acknowledged for providing the Al--Mn alloy.


\bibliographystyle{abbrvnat}
\bibliography{library_sanitized}

@Article{wright2015electron,
  author    = {Wright, Stuart I. and Nowell, Matthew M. and {De Kloe}, Ren{\'e} and Camus, Patrick and Rampton, Travis},
  title     = {{Electron imaging with an EBSD detector}},
  doi       = {10.1016/j.ultramic.2014.10.002},
  pages     = {132--145},
  volume    = {148},
  journal   = {Ultramicroscopy},
  publisher = {Elsevier},
  year      = {2015},
}

@Article{bachmann2011grain,
  author    = {Bachmann, Florian and Hielscher, Ralf and Schaeben, Helmut},
  title     = {{Grain detection from 2d and 3d EBSD data---Specification of the MTEX algorithm}},
  doi       = {10.1016/j.ultramic.2011.08.002},
  number    = {12},
  pages     = {1720--1733},
  volume    = {111},
  journal   = {Ultramicroscopy},
  publisher = {Elsevier},
  year      = {2011},
}

@Article{chen2015dictionary,
  author   = {Chen, Yu H. and Park, Se Un and Wei, Dennis and Newstadt, Greg and Jackson, Michael A. and Simmons, Jeff P. and {De Graef}, Marc and Hero, Alfred O.},
  title    = {{A Dictionary Approach to Electron Backscatter Diffraction Indexing}},
  doi      = {10.1017/S1431927615000756},
  number   = {3},
  pages    = {739--752},
  volume   = {21},
  journal  = {Microscopy and Microanalysis},
  year     = {2015},
}

@Article{wright2015introduction,
  author    = {Wright, Stuart I. and Nowell, Matthew M. and Lindeman, Scott P. and Camus, Patrick P. and {De Graef}, Marc and Jackson, Michael A.},
  title     = {{Introduction and comparison of new EBSD post-processing methodologies}},
  doi       = {10.1016/j.ultramic.2015.08.001},
  pages     = {81--94},
  volume    = {159},
  journal   = {Ultramicroscopy},
  publisher = {Elsevier},
  year      = {2015},
}

@Article{ram2015error,
  author     = {Ram, Farangis and Zaefferer, Stefan and J{\"a}pel, Tom and Raabe, Dierk},
  title      = {{Error analysis of the crystal orientations and disorientations obtained by the classical electron backscatter diffraction technique}},
  doi        = {10.1107/S1600576715005762},
  pages      = {797--813},
  volume     = {48},
  journal    = {Journal of Applied Crystallography},
  year       = {2015},
}

@Article{krakow2017onthree,
  author   = {Krakow, Robert and Bennett, Robbie J and Johnstone, Duncan N and Vukmanovic, Zoja and Solano-Alvarez, Wilberth and Lain{\'e}, Steven J and Einsle, Joshua F and Midgley, Paul A and Rae, Catherine M F and Hielscher, Ralf},
  title    = {{On three-dimensional misorientation spaces}},
  doi      = {10.1098/rspa.2017.0274},
  number   = {2206},
  pages    = {20170274},
  volume   = {473},
  journal  = {Proceedings of the Royal Society A},
  pmid     = {29118660},
  year     = {2017},
}

@Article{nolze2017electron,
  author   = {Nolze, Gert and Hielscher, Ralf and Winkelmann, Aimo},
  title    = {{Electron backscatter diffraction beyond the mainstream}},
  doi      = {10.1002/crat.201600252},
  number   = {1},
  pages    = {1--24},
  volume   = {52},
  journal  = {Crystal Research and Technology},
  year     = {2017},
}

@article{hielscher2008novel,
	author = {Hielscher, R. and Schaeben, H.},
	doi = {10.1107/S0021889808030112},
	journal = {Journal of Applied Crystallography},
	number = {6},
	pages = {1024--1037},
	publisher = {International Union of Crystallography},
	title = {{A novel pole figure inversion method: Specification of the MTEX algorithm}},
	volume = {41},
	year = {2008}}

@article{makhlouf2001aluminum,
	author = {Makhlouf, M. M. and Guthy, H. V.},
	doi = {10.1016/S1471-5317(02)00003-2},
	journal = {Journal of Light Metals},
	number = {4},
	pages = {199--218},
	title = {{The aluminum-silicon eutectic reaction: Mechanisms and crystallography}},
	volume = {1},
	year = {2001}}

@Article{britton2018astro,
  author        = {Britton, Thomas Benjamin and Tong, Vivian and Hickey, Jim and Foden, Alex and Wilkinson, Angus},
  title         = {{AstroEBSD: exploring new space in pattern indexing with methods launched from an astronomical approach}},
  doi           = {10.1107/S1600576718010373},
  pages         = {1--10},
  volume        = {51},
  journal       = {Journal of Applied Crystallography},
  publisher     = {International Union of Crystallography},
  year          = {2018},
}

@Article{callahan2013dynamical,
  author        = {Callahan, Patrick G and {De Graef}, Marc},
  title         = {{Dynamical Electron Backscatter Diffraction Patterns. Part I: Pattern Simulations}},
  doi           = {10.1017/S1431927613001840},
  pages         = {1255--1265},
  volume        = {19},
  isbn          = {1431927613001},
  journal       = {Microscopy and Microanalysis},
  year          = {2013},
}

@article{singh2016orientation,
	author = {Singh, S. and {De Graef}, Marc},
	doi = {10.1088/0965-0393/24/8/085013},
	journal = {Modelling and Simulation in Materials Science and Engineering},
	number = {8},
	publisher = {IOP Publishing},
	title = {{Orientation sampling for dictionary-based diffraction pattern indexing methods}},
	volume = {24},
	year = {2016}}

@Article{britton2016tutorial,
  author    = {Britton, T. B. and Jiang, J. and Guo, Y. and Vilalta-Clemente, A. and Wallis, D. and Hansen, L. N. and Winkelmann, A. and Wilkinson, A. J.},
  title     = {{Tutorial: Crystal orientations and EBSD - Or which way is up?}},
  doi       = {10.1016/j.matchar.2016.04.008},
  pages     = {113--126},
  volume    = {117},
  journal   = {Materials Characterization},
  publisher = {The Authors},
  year      = {2016},
}

@Book{schwartz2009electron,
  author    = {Schwartz, Adam J and Kumar, Mukul and Adams, Brent L and Field, David P},
  title     = {{Electron Backscatter Diffraction in Materials Science}},
  doi       = {10.1007/978-0-387-88136-2},
  edition   = {2nd},
  publisher = {Springer},
  ed        = {2nd},
  year      = {2009},
}

@Article{marquardt2017quantitative,
  author   = {Marquardt, Katharina and {De Graef}, Marc and Singh, Saransh and Marquardt, Hauke and Rosenthal, Anja and Koizuimi, Sanae},
  title    = {{Quantitative electron backscatter diffraction (EBSD) data analyses using the dictionary indexing (DI) approach: Overcoming indexing difficulties on geological materials}},
  doi      = {10.2138/am-2017-6062},
  number   = {9},
  pages    = {1843--1855},
  volume   = {102},
  journal  = {American Mineralogist},
  year     = {2017},
}

@InProceedings{schwarzer2011imaging,
  author    = {Schwarzer, Robert and Sukkau, Johann and Hjelen, Jarle},
  booktitle = {Microscopy Conference Kiel},
  title     = {{Imaging of topography and phase distributions with an EBSD detector in the SEM}},
  year      = {2011},
}

@Article{ram2017error,
  author     = {Ram, Farangis and Wright, Stuart and Singh, Saransh and {De Graef}, Marc},
  title      = {{Error analysis of the crystal orientations obtained by the dictionary approach to EBSD indexing}},
  doi        = {10.1016/j.ultramic.2017.04.016},
  pages      = {17--26},
  volume     = {181},
  journal    = {Ultramicroscopy},
  publisher  = {Elsevier},
  year       = {2017},
}

@article{singh2018high,
	author = {Singh, Saransh and Guo, Yi and Winiarski, Bart{\l}omiej and Burnett, Timothy L and Withers, Philip J and {De Graef}, Marc},
	doi = {10.1038/s41598-018-29315-8},
	journal = {Scientific reports},
	publisher = {Nature Publishing Group},
	title = {{High resolution low kV EBSD of heavily deformed and nanocrystalline Aluminium by dictionary-based indexing}},
	volume = {8},
	year = {2018}}

@Article{jackson2019dictionary,
  author    = {Jackson, MA and Pascal, E and {De Graef}, M},
  title     = {{Dictionary Indexing of Electron Back-Scatter Diffraction Patterns: a Hands-On Tutorial}},
  doi       = {10.1007/s40192-019-00137-4},
  pages     = {1--21},
  volume    = {8},
  journal   = {Integrating Materials and Manufacturing Innovation},
  publisher = {Springer},
  year      = {2019},
}

@inproceedings{winkelmann2016physics,
	author = {Winkelmann, Aimo and Nolze, Gert and Vos, Maarten and Salvat-Pujol, Francesc and Werner, WSM},
	booktitle = {{IOP Conference Series: Materials Science and Engineering}},
	doi = {10.1088/1757-899X/109/1/012018},
	number = {1},
	organization = {IOP Publishing},
	pages = {012018},
	title = {{Physics-based simulation models for EBSD: advances and challenges}},
	volume = {109},
	year = {2016}}

@Article{groeber2014dream,
  author     = {Groeber, Michael A and Jackson, Michael A},
  title      = {{DREAM.3D: a digital representation environment for the analysis of microstructure in 3D}},
  doi        = {10.1186/2193-9772-3-5},
  number     = {1},
  pages      = {5},
  volume     = {3},
  journal    = {Integrating Materials and Manufacturing Innovation},
  publisher  = {Springer},
  year       = {2014},
}

@article{pena2017electron,
	author = {de la Pe{\~n}a, Francisco and Ostasevicius, Tomas and Fauske, Vidar Tonaas and Burdet, Pierre and Jokubauskas, Petras and Nord, Magnus and Sarahan, Mike and Prestat, Eric and Johnstone, Duncan N and Taillon, Joshua and others},
	doi = {10.1017/S1431927617001751},
	journal = {Microscopy and Microanalysis},
	number = {S1},
	pages = {214--215},
	publisher = {Cambridge University Press},
	title = {{Electron microscopy (Big and Small) data analysis with the open source software package HyperSpy}},
	volume = {23},
	year = {2017}}

@article{winkelmann2007many,
	author = {Winkelmann, Aimo and Trager-Cowan, Carol and Sweeney, Francis and Day, Austin P and Parbrook, Peter},
	doi = {10.1016/j.ultramic.2006.10.006},
	journal = {Ultramicroscopy},
	number = {4-5},
	pages = {414--421},
	publisher = {Elsevier},
	title = {{Many-beam dynamical simulation of electron backscatter diffraction patterns}},
	volume = {107},
	year = {2007}}

@book{gonzalez2017digital,
	author = {Gonzalez, Rafael C and Woods, Richard E},
	edition = {4th},
	isbn = {978-0133356724},
	publisher = {Pearson Education Limited},
	title = {{Digital Image Processing}},
	year = {2017}}

@article{zhu2019automated,
	author = {Zhu, Chaoyi and Kaufmann, Kevin and Vecchio, Kenneth},
	doi = {10.1017/S1431927619000710},
	journal = {Microscopy and Microanalysis},
	pages = {1--12},
	publisher = {Cambridge University Press},
	title = {{Automated Reconstruction of Spherical Kikuchi Maps}},
	year = {2019}}

@Article{cooper1966crystal,
  author    = {Cooper, Malcolm and Robinson, K},
  title     = {{The Crystal Structure of the Ternary Alloy $\alpha$(AlMnSi)}},
  doi       = {10.1107/S0365110X6600149X},
  number    = {5},
  pages     = {614--617},
  volume    = {20},
  journal   = {Acta Crystallographica},
  publisher = {International Union of Crystallography},
  year      = {1966},
}

@article{bordin2016ebsd,
	author = {Bord{\'\i}n, S Fernandez and Limandri, S and Ranalli, JM and Castellano, G},
	doi = {10.1016/j.ultramic.2016.09.010},
	journal = {Ultramicroscopy},
	pages = {177--185},
	publisher = {Elsevier},
	title = {{EBSD spatial resolution for detecting sigma phase in steels}},
	volume = {171},
	year = {2016}}

@Article{lenthe2019spherical,
  author    = {Lenthe, WC and Singh, S and {De Graef}, M},
  title     = {{A spherical harmonic transform approach to the indexing of electron back-scattered diffraction patterns}},
  doi       = {10.1016/j.ultramic.2019.112841},
  pages     = {112841},
  volume    = {207},
  journal   = {Ultramicroscopy},
  publisher = {Elsevier},
  year      = {2019},
}

@Article{singh2017application,
  author     = {Singh, Saransh and Ram, Farangis and {De Graef}, Marc},
  title      = {{Application of forward models to crystal orientation refinement}},
  doi        = {10.1107/S1600576717014200},
  number     = {6},
  pages      = {1664--1676},
  volume     = {50},
  journal    = {Journal of Applied Crystallography},
  publisher  = {International Union of Crystallography},
  year       = {2017},
}

@article{kasper1954ordering,
	author = {Kasper, JS},
	doi = {10.1016/0001-6160(54)90066-8},
	journal = {Acta Metallurgica},
	number = {3},
	pages = {456--461},
	publisher = {Elsevier},
	title = {{The ordering of atoms in the chi-phase of the iron-chromium-molybdenum system}},
	volume = {2},
	year = {1954}}

@Article{ram2018phase,
  author     = {Ram, Farangis and {De Graef}, Marc},
  title      = {{Phase differentiation by electron backscatter diffraction using the dictionary indexing approach}},
  doi        = {10.1016/j.actamat.2017.10.069},
  pages      = {352--364},
  volume     = {144},
  journal    = {Acta Materialia},
  publisher  = {Elsevier},
  year       = {2018},
}

@PhdThesis{lassen1994automated,
  author = {Lassen, Niels Christian Krieger},
  title  = {{Automated Determination of Crystal Orientations from Electron Backscattering Patterns}},
  url    = {http://ebsd.info/pdf/PhD_KriegerLassen.pdf},
  school = {Institute of Mathematical Modelling},
  year   = {1994},
}

@article{johnstone2020density,
	author = {Johnstone, Duncan N and Martineau, Ben H and Crout, Phillip and Midgley, Paul A and Eggeman, Alexander S},
	doi = {10.1107/S1600576720011103},
	journal = {Journal of Applied Crystallography},
	number = {5},
	publisher = {International Union of Crystallography},
	title = {{Density-based clustering of crystal (mis)orientations and the orix Python library}},
	volume = {53},
	year = {2020}}

@Article{kontio1981new,
  author    = {Kontio, A and Coppens, Pv},
  title     = {{New study of the structure of MnAl$_6$}},
  doi       = {10.1107/S0567740881003191},
  number    = {2},
  pages     = {433--435},
  volume    = {37},
  journal   = {Acta Crystallographica Section B: Structural Crystallography and Crystal Chemistry},
  publisher = {International Union of Crystallography},
  year      = {1981},
}

@techreport{sjalander2019epic,
	archiveprefix = {arXiv},
	author = {Sj{\"a}lander, Magnus and Jahre, Magnus and Tufte, Gunnar and Reissmann, Nico},
	doi = {10.48550/arXiv.1912.05848},
	eprint = {1912.05848},
	primaryclass = {cs.DC},
	title = {{{EPIC}: An Energy-Efficient, High-Performance {GPGPU} Computing Research Infrastructure}},
	year = {2019}}

@article{winkelmann2020refined,
	author = {Winkelmann, Aimo and Nolze, Gert and Cios, Grzegorz and Tokarski, Tomasz and Ba{\l}a, Piotr},
	doi = {10.3390/ma13122816},
	journal = {Materials},
	number = {12},
	pages = {2816},
	publisher = {Multidisciplinary Digital Publishing Institute},
	title = {{Refined Calibration Model for Improving the Orientation Precision of Electron Backscatter Diffraction Maps}},
	volume = {13},
	year = {2020}}

@inproceedings{aanes2020processing,
	author = {{\AA}nes, H. W. and Hjelen, J. and S{\o}rensen, B. E. and van Helvoort, A. T. J. and Marthinsen, K.},
	booktitle = {{IOP Conference Series: Materials Science and Engineering}},
	doi = {10.1088/1757-899X/891/1/012002},
	number = {1},
	organization = {IOP Publishing},
	pages = {012002},
	title = {{Processing and indexing of electron backscatter patterns using open-source software}},
	volume = {891},
	year = {2020}}

@InProceedings{kluyver2016jupyter,
  author    = {Kluyver, Thomas and Ragan-Kelley, Benjamin and P{\'e}rez, Fernando and Granger, Brian E and Bussonnier, Matthias and Frederic, Jonathan and Kelley, Kyle and Hamrick, Jessica B and Grout, Jason and Corlay, Sylvain and others},
  booktitle = {{Positioning and Power in Academic Publishing: Players, Agents and Agendas}},
  title     = {{Jupyter Notebooks---a publishing format for reproducible computational workflows.}},
  doi       = {10.3233/978-1-61499-649-1-87},
  pages     = {87--90},
  year      = {2016},
}

@Article{nolze2016pattern,
  author    = {Nolze, Gert and Winkelmann, Aimo and Boyle, Alan P},
  title     = {{Pattern matching approach to pseudosymmetry problems in electron backscatter diffraction}},
  doi       = {10.1016/j.ultramic.2015.10.010},
  pages     = {146--154},
  volume    = {160},
  journal   = {Ultramicroscopy},
  publisher = {Elsevier},
  year      = {2016},
}

@article{kiesheyer1976investigation,
	author = {Kiesheyer, H and Brandis, H},
	journal = {Zeitschrift fuer Metallkunde},
	number = {4},
	pages = {258--263},
	title = {{Investigation of phase equilibria in the ternary system Fe-Cr-Mo in solid state}},
	volume = {67},
	year = {1976}}

@book{wyckoff1963crystal,
	author = {Wyckoff, Ralph Walter Graystone},
	publisher = {Interscience publishers New York},
	title = {{Crystal structures}},
	volume = {1},
	year = {1963}}

@article{virtanen2020scipy,
	adsurl = {https://rdcu.be/b08Wh},
	author = {Virtanen, Pauli and Gommers, Ralf and Oliphant, Travis E. and Haberland, Matt and Reddy, Tyler and Cournapeau, David and Burovski, Evgeni and Peterson, Pearu and Weckesser, Warren and Bright, Jonathan and {van der Walt}, St{\'e}fan J. and Brett, Matthew and Wilson, Joshua and Millman, K. Jarrod and Mayorov, Nikolay and Nelson, Andrew R. J. and Jones, Eric and Kern, Robert and Larson, Eric and Carey, C J and Polat, Ilhan and Feng, Yu and Moore, Eric W. and {VanderPlas}, Jake and Laxalde, Denis and Perktold, Josef and Cimrman, Robert and Henriksen, Ian and Quintero, E. A. and Harris, Charles R. and Archibald, Anne M. and Ribeiro, Antonio H. and Pedregosa, Fabian and {van Mulbregt}, Paul and {SciPy 1.0 Contributors}},
	doi = {10.1038/s41592-019-0686-2},
	journal = {Nature Methods},
	pages = {261--272},
	title = {{{SciPy} 1.0: Fundamental Algorithms for Scientific Computing in Python}},
	volume = {17},
	year = {2020}}

@article{wilkinson2016fair,
	author = {Wilkinson, Mark D and Dumontier, Michel and Aalbersberg, IJsbrand Jan and Appleton, Gabrielle and Axton, Myles and Baak, Arie and Blomberg, Niklas and Boiten, Jan-Willem and da Silva Santos, Luiz Bonino and Bourne, Philip E and others},
	doi = {10.1038/sdata.2016.18},
	journal = {Scientific data},
	number = {1},
	pages = {1--9},
	publisher = {Nature Publishing Group},
	title = {{The FAIR Guiding Principles for scientific data management and stewardship}},
	volume = {3},
	year = {2016}}

@Article{aanes2022correlated,
  author     = {{\AA}nes, H{\aa}kon W. and {van Helvoort}, Antonius T. J. and Marthinsen, Knut},
  title      = {{Correlated subgrain and particle analysis of a recovered Al-Mn alloy by directly combining EBSD and backscatter electron imaging}},
  doi        = {10.1016/j.matchar.2022.112228},
  pages      = {112228},
  volume     = {192},
  journal    = {Materials Characterization},
  year       = {2022},
}

@Article{schultheiss2022confinement,
  author   = {Schulthei{\ss}, Jan and Xue, Fei and Roede, Erik and {\AA}nes, H{\aa}kon W. and Danmo, Frida H. and Selbach, Sverre M. and Chen, Long-Qing and Meier, Dennis},
  title    = {{Confinement-Driven Inverse Domain Scaling in Polycrystalline ErMnO$_3$}},
  doi      = {10.1002/adma.202203449},
  issue    = {45},
  pages    = {2203449},
  volume   = {34},
  journal  = {Advanced Materials},
  year     = {2022},
}

@article{akselsen2021effect,
	author = {Akselsen, Odd M and Bj{\o}rge, Ruben and {\AA}nes, H{\aa}kon Wiik and Ren, Xiaobo and Nyhus, B{\aa}rd},
	doi = {10.3390/met11122045},
	journal = {Metals},
	number = {12},
	pages = {2045},
	publisher = {MDPI},
	title = {{Effect of Sigma Phase in Wire Arc Additive Manufacturing of Superduplex Stainless Steel}},
	volume = {11},
	year = {2021}}

@article{akselsen2022microstructure,
	author = {Akselsen, Odd M. and Bj{\o}rge, Ruben and {\AA}nes, H{\aa}kon Wiik and Ren, Xiaobo and Nyhus, B{\aa}rd},
	doi = {10.3390/met12111867},
	journal = {Metals},
	number = {11},
	pages = {1867},
	title = {{Microstructure and Properties of Wire Arc Additive Manufacturing of Inconel 625}},
	volume = {12},
	year = {2022}}

@article{bergh2023intermetallic,
	author = {Bergh, Tina and {\AA}nes, H{\aa}kon Wiik and Aune, Ragnhild and Wenner, Sigurd and Holmestad, Randi and Ren, Xiaobo and Vullum, Per Erik},
	doi = {10.2320/matertrans.MT-LA2022046},
	journal = {Materials Transactions},
	number = {2},
	pages = {352-359},
	title = {{Intermetallic Phase Layers in Cold Metal Transfer Aluminium-Steel Welds with an Al--Si--Mn Filler Alloy}},
	volume = {64},
	year = {2023}}

@article{morawiec1996rodrigues,
	author = {Morawiec, A and Field, DP},
	doi = {10.1080/01418619608243708},
	journal = {Philosophical Magazine A},
	number = {4},
	pages = {1113--1130},
	publisher = {Taylor \& Francis},
	title = {Rodrigues parameterization for orientation and misorientation distributions},
	volume = {73},
	year = {1996}}

@article{liu2015twin,
	author = {Liu, Xiaorui and Zhang, Yudong and Beausir, Beno{\^\i}t and Liu, Fang and Esling, Claude and Yu, Fuxiao and Zhao, Xiang and Zuo, Liang},
	doi = {10.1016/j.actamat.2015.06.041},
	journal = {Acta Materialia},
	pages = {338--347},
	publisher = {Elsevier},
	title = {{Twin-controlled growth of eutectic Si in unmodified and Sr-modified Al--12.7\% Si alloys investigated by SEM/EBSD}},
	volume = {97},
	year = {2015}}

@Article{nowell2004phase,
  author    = {Nowell, Matt M and Wright, Stuart I},
  title     = {{Phase differentiation via combined EBSD and XEDS}},
  doi       = {10.1111/j.0022-2720.2004.01299.x},
  number    = {3},
  pages     = {296--305},
  volume    = {213},
  journal   = {Journal of microscopy},
  publisher = {Wiley Online Library},
  year      = {2004},
}

@article{shamsuzzoha1986crystal,
	author = {Shamsuzzoha, M and Hogan, LM},
	doi = {10.1080/01418618608243605},
	journal = {Philosophical Magazine A},
	number = {4},
	pages = {459--477},
	publisher = {Taylor \& Francis},
	title = {{The crystal morphology of fibrous silicon in strontium-modified Al-Si eutectic}},
	volume = {54},
	year = {1986}}

@Article{aanes2023orientation,
  author     = {{\AA}nes, H{\aa}kon W. and {van Helvoort}, Antonius T. J. and Marthinsen, Knut},
  title      = {Orientation dependent pinning of (sub)grains by dispersoids during recovery and recrystallization in an Al-Mn alloy},
  doi        = {10.1016/j.actamat.2023.118761},
  pages      = {118761},
  volume     = {248},
  journal    = {Acta Materialia},
  publisher  = {Elsevier},
  year       = {2023},
}

@Article{day1968microstructure,
  author    = {Day, MG and Hellawell, A},
  title     = {{The microstructure and crystallography of aluminium---silicon eutectic alloys}},
  doi       = {10.1098/rspa.1968.0128},
  number    = {1483},
  pages     = {473--491},
  volume    = {305},
  journal   = {Proceedings of the Royal Society of London. Series A. Mathematical and Physical Sciences},
  publisher = {The Royal Society London},
  year      = {1968},
}

@Book{prince2004international,
  title       = {International Tables for Crystallography, Volume C: Mathematical, physical and chemical tables},
  doi         = {10.1107/97809553602060000103},
  editor      = {Prince, Edward},
  publisher   = {Springer Science \& Business Media},
  year        = {2006},
}

@Software{hyperspy_concept_doi,
  author       = {de la Pe{\~n}a, Francisco and Prestat, Eric and Fauske, Vidar Tonaas and Burdet, Pierre and L{\"a}hnemann, Jonas and Jokubauskas, Petras and Furnival, Tom and Nord, Magnus and Ostasevicius, Tomas and MacArthur, Katherine E. and Johnstone, Duncan N. and Sarahan, Mike and Taillon, Joshua and Aarholt, Thomas and pquinn-dls and Migunov, Vadim and Eljarrat, Alberto and Caron, Jan and Francis, Carter and Nemoto, T. and Poon, Timothy and Mazzucco, Stefano and actions-user and Tappy, Nicolas and Cautaerts, Niels and Somnath, Suhas and Slater, Tom and Walls, Michael and Winkler, Florian and {\AA}nes, H{\aa}kon Wiik},
  title        = {hyperspy},
  doi          = {10.5281/zenodo.592838},
  howpublished = {\url{https://doi.org/10.5281/zenodo.592838}},
  note         = {Zenodo},
  publisher    = {Zenodo},
  year         = {2023},
}

@Article{nishihara2012isothermal,
  author    = {Nishihara, Yu and Nakajima, Yoichi and Akashi, Akihiko and Tsujino, Noriyoshi and Takahashi, Eiichi and Funakoshi, Ken-ichi and Higo, Yuji},
  title     = {Isothermal compression of face-centered cubic iron},
  doi       = {10.2138/am.2012.3958},
  number    = {8-9},
  pages     = {1417--1420},
  volume    = {97},
  journal   = {American Mineralogist},
  publisher = {Mineralogical Society of America},
  year      = {2012},
}

@Article{nenno1962orientation,
  author     = {Nenno, Soji and Tagaya, Masayoshi and Nishiyama, Zenji},
  title      = {{Orientation Relationships between Gamma (f.c.c.) and Sigma Phases in an Iron-Chromium-Nickel Alloy}},
  doi        = {10.2320/matertrans1960.3.82},
  number     = {2},
  pages      = {82--93},
  volume     = {3},
  journal    = {Transactions of the Japan Institute of Metals},
  publisher  = {The Japan Institute of Metals},
  year       = {1962},
}

@Article{chen2001effects,
  author    = {Chen, TH and Yang, JR},
  title     = {{Effects of solution treatment and continuous cooling on $\sigma$-phase precipitation in a 2205 duplex stainless steel}},
  doi       = {10.1016/S0921-5093(01)00911-X},
  number    = {1-2},
  pages     = {28--41},
  volume    = {311},
  journal   = {Materials Science and Engineering: A},
  publisher = {Elsevier},
  year      = {2001},
}

@article{lee2012isothermal,
	author = {Lee, Tae-Ho and Ha, Heon-Young and Hwang, Byoungchul and Kim, Sung-Joon},
	doi = {10.1007/s11661-011-0935-1},
	journal = {Metallurgical and Materials Transactions A},
	number = {3},
	pages = {822--832},
	publisher = {Springer},
	title = {Isothermal decomposition of ferrite in a high-nitrogen, nickel-free duplex stainless steel},
	volume = {43},
	year = {2012}}

@article{redjaimia2004morphology,
	author = {Redja{\"\i}mia, A and Proult, A and Donnadieu, P and Morniroli, JP},
	doi = {10.1023/B:JMSC.0000019999.27065.13},
	journal = {Journal of materials science},
	number = {7},
	pages = {2371--2386},
	publisher = {Springer},
	title = {Morphology, crystallography and defects of the intermetallic $\chi$-phase precipitated in a duplex ($\delta$+ $\gamma$) stainless steel},
	volume = {39},
	year = {2004}}

@Article{nolze2004characterization,
  author     = {Nolze, Gert},
  title      = {{Characterization of the fcc/bcc orientation relationship by EBSD using pole figures and variants}},
  doi        = {10.1515/ijmr-2004-0142},
  number     = {9},
  pages      = {744--755},
  volume     = {95},
  journal    = {International Journal of Materials Research},
  publisher  = {De Gruyter},
  year       = {2004},
}

@Article{lee2005crystal,
  author   = {Lee, Tae-Ho and Oh, Chang-Seok and Han, Heung Nam and Lee, Chang Gil and Kim, Sung-Joon and Takaki, Setsuo},
  title    = {{On the crystal structure of Cr$_2$N precipitates in high-nitrogen austenitic stainless steel}},
  doi      = {10.1107/S0108768104033919},
  number   = {2},
  pages    = {137-144},
  volume   = {61},
  journal  = {Acta Crystallographica Section B},
  year     = {2005},
}

@Article{ramirez2003relationship,
  author  = {Ramirez, AJ and Lippold, JC and Brandi, SD},
  title   = {{The Relationship between Chromium Nitride and Secondary Austenite Precipitation in Duplex Stainless Steels}},
  doi     = {10.1007/s11661-003-0304-9},
  number  = {8},
  pages   = {1575--1597},
  volume  = {34},
  journal = {Metallurgical and materials transactions A},
  year    = {2003},
}

@Article{pettersson2015precipitation,
  author    = {Pettersson, Niklas and Pettersson, Rachel FA and Wessman, Sten},
  title     = {{Precipitation of chromium nitrides in the super duplex stainless steel 2507}},
  doi       = {10.1007/s11661-014-2718-y},
  pages     = {1062--1072},
  volume    = {46},
  journal   = {Metallurgical and Materials Transactions A},
  publisher = {Springer},
  year      = {2015},
}

@Article{nilsson1992super,
  author    = {Nilsson, J.-O.},
  title     = {Super duplex stainless steels},
  doi       = {10.1179/mst.1992.8.8.685},
  number    = {8},
  pages     = {685-700},
  volume    = {8},
  journal   = {Materials Science and Technology},
  publisher = {Taylor & Francis},
  year      = {1992},
}

@article{liu1990metastable,
	author = {Liu, S and Hamagochi, Y and Kuwano, H},
	journal = {Acta Metallurgica Sinica(China)},
	number = {6},
	title = {{Metastable R Phase in Fe--Cr--Mo and Fe--Cr--W Alloys}},
	volume = {26},
	year = {1990}}

@article{shoemaker1978refinement,
	author = {Shoemaker, C. B. and Shoemaker, D. P. and Hopkins, T. E. and Yindepit, S.},
	doi = {10.1107/S0567740878011620},
	journal = {Acta Crystallographica Section B},
	number = {12},
	pages = {3573--3576},
	title = {{Refinement of the structure of {$\beta$}-manganese and of a related phase in the Mn{--}Ni{--}Si system}},
	volume = {34},
	year = {1978}}

@Article{bugten2023role,
  author    = {Bugten, AV and Michels, L and Brurok, RB and Hartung, C and Ott, E and Vines, L and Li, Y and Arnberg, L and Di Sabatino, M},
  title     = {{The Role of Boron in Low Copper Spheroidal Graphite Irons}},
  doi       = {10.1007/s11661-023-07014-y},
  pages     = {1--15},
  journal   = {Metallurgical and Materials Transactions A},
  publisher = {Springer},
  year      = {2023},
}

@article{donnadieu1994manganese,
	author = {Donnadieu, P and Lapasset, G and Sanders, TH},
	doi = {10.1080/09500839408240993},
	journal = {Philosophical Magazine Letters},
	number = {5},
	pages = {319--326},
	publisher = {Taylor \& Francis},
	title = {{Manganese-induced ordering in the $\alpha$-(Al-Mn-Fe-Si) approximant phase}},
	volume = {70},
	year = {1994}}

@Article{nolze2015kikuchi,
  author   = {Nolze, G. and Grosse, C. and Winkelmann, A.},
  title    = {Kikuchi pattern analysis of noncentrosymmetric crystals},
  doi      = {10.1107/S1600576715014016},
  number   = {5},
  pages    = {1405-1419},
  volume   = {48},
  journal  = {Journal of Applied Crystallography},
  year     = {2015},
}

@inproceedings{rocklin2015dask,
	author = {Rocklin, Matthew},
	booktitle = {Proceedings of the 14th python in science conference},
	organization = {SciPy Austin, TX},
	pages = {136},
	title = {Dask: Parallel computation with blocked algorithms and task scheduling},
	volume = {130},
	year = {2015}}

@InProceedings{lam2015numba,
  author    = {Lam, Siu Kwan and Pitrou, Antoine and Seibert, Stanley},
  booktitle = {Proceedings of the Second Workshop on the LLVM Compiler Infrastructure in HPC},
  title     = {{Numba: A LLVM-Based Python JIT Compiler}},
  doi       = {10.1145/2833157.2833162},
  location  = {Austin, Texas},
  publisher = {Association for Computing Machinery},
  series    = {LLVM '15},
  address   = {New York, NY, USA},
  articleno = {7},
  numpages  = {6},
  year      = {2015},
}

@Software{orix_concept_doi,
  author       = {{\AA}nes, H{\aa}kon Wiik and Martineau, Ben and Harrison, Paddy and Crout, Phillip and Johnstone, Duncan and Cautaerts, Niels and Gerlt, Austin and Mathisen, Anders Christian and H{\o}g{\aa}s, Simon and Clausen, Alexander},
  title        = {orix},
  doi          = {10.5281/zenodo.3459662},
  howpublished = {\url{https://doi.org/10.5281/zenodo.3459662}},
  note         = {Zenodo},
  journal      = {Zenodo},
  year         = {2023},
}

@Software{diffsims_concept_doi,
  author       = {Johnstone, Duncan and Crout, Phillip and {\AA}nes, H{\aa}kon Wiik and Prestat, Eric and Tovey, Rob and H{\o}g{\aa}s, Simon and Martineau, Ben and Laulainen, Joonatan and Cautaerts, Niels and Wood, Isabel and Collins, Sean and Smeets, Stef and Borrelli, Alex and Doherty, Tiarnan and Morzy, Jedrzej and Jacobsen, Endre and Bergh, Tina and Ostasevicius, Tomas and Opheim, Eirik},
  title        = {diffsims},
  doi          = {10.5281/zenodo.3337900},
  howpublished = {\url{https://doi.org/10.5281/zenodo.3337900}},
  note         = {Zenodo},
  publisher    = {Zenodo},
  year         = {2023},
}

@Article{lassen1992image,
  author    = {Lassen, NC Krieger and Jensen, D Juul and Conradsen, KJSM},
  title     = {{Image Processing Procedures for Analysis of Electron Back Scattering Patterns}},
  pages     = {115--121},
  volume    = {6},
  journal   = {Scanning microscopy},
  publisher = {Scanning Microscopy International},
  year      = {1992},
}

@Software{nlopt,
  author       = {Johnson, Steven G.},
  title        = {The {NLopt} nonlinear-optimization package},
  howpublished = {\url{https://github.com/stevengj/nlopt}},
  year         = {2007},
}

@Article{sandvik2023pressure,
  author  = {Sandvik, O W and M{\"u}ller, A M and {\AA}nes, H W and Zahn, M and He, J and Fiebig, M and Lottermoser, Th and Rojac, T and Meier, D and Schulthei{\ss}, J},
  title   = {{Pressure Control of Nonferroelastic Ferroelectric Domains in ErMnO$_3$}},
  doi     = {10.1021/acs.nanolett.3c01638},
  number  = {15},
  pages   = {6994-7000},
  volume  = {23},
  journal = {Nano Letters},
  year    = {2023},
}

@Article{cios2023mapping,
  author   = {Cios, Grzegorz and Winkelmann, Aimo and Nolze, Gert and Tokarski, Tomasz and Rych{\l}owski, {\L}ukasz and Dan, Leonid and Ba{\l}a, Piotr},
  title    = {{Mapping of lattice distortion in martensitic steel---Comparison of different evaluation methods of EBSD patterns}},
  doi      = {10.1016/j.ultramic.2023.113824},
  issn     = {0304-3991},
  pages    = {113824},
  journal  = {Ultramicroscopy},
  year     = {2023},
}

@Article{zemanek1983algorithmic,
  author     = {Zemanek, Heinz},
  title      = {Algorithmic Perfection},
  doi        = {10.1109/MAHC.1983.10000},
  number     = {1},
  pages      = {73-73},
  volume     = {5},
  journal    = {Annals of the History of Computing},
  year       = {1983},
}

@Article{okazaki2008effects,
  author     = {Okazaki, Yoshimitsu},
  title      = {{Effects of heat treatment and hot forging on microstructure and mechanical properties of Co-Cr-Mo alloy for surgical implants}},
  doi        = {10.2320/matertrans.MRA2007274},
  number     = {4},
  pages      = {817--823},
  volume     = {49},
  journal    = {Materials transactions},
  publisher  = {The Japan Institute of Metals and Materials},
  year       = {2008},
}

@article{pinard2011open,
	author = {Pinard, Philippe T and Lagac{\'e}, Marin and Hovington, Pierre and Thibault, Denis and Gauvin, Raynald},
	doi = {10.1017/S1431927611000456},
	journal = {Microscopy and Microanalysis},
	number = {3},
	pages = {374--385},
	publisher = {Cambridge University Press},
	title = {{An open-source engine for the processing of electron backscatter patterns: EBSD-Image}},
	volume = {17},
	year = {2011}}

@Article{rowenhorst2024fast,
  author     = {Rowenhorst, David J and Callahan, Patrick G and {\AA}nes, H{\aa}kon W},
  title      = {{Fast Radon transforms for high-precision EBSD orientation determination using PyEBSDIndex}},
  doi        = {10.1107/S1600576723010221},
  number     = {1},
  volume     = {57},
  journal    = {Journal of Applied Crystallography},
  publisher  = {International Union of Crystallography},
  year       = {2024},
}

@Article{wright1992automatic,
  author    = {Wright, Stuart I and Adams, Brent L},
  title     = {Automatic analysis of electron backscatter diffraction patterns},
  doi       = {10.1007/BF02675553},
  number    = {3},
  pages     = {759--767},
  volume    = {23},
  journal   = {Metallurgical Transactions A},
  publisher = {Springer},
  year      = {1992},
}

@Article{nilsson1993influence,
  author    = {Nilsson, JO and Wilson, A},
  title     = {{Influence of isothermal phase transformations on toughness and pitting corrosion of super duplex stainless steel SAF 2507}},
  doi       = {10.1179/mst.1993.9.7.545},
  number    = {7},
  pages     = {545--554},
  volume    = {9},
  journal   = {Materials Science and Technology},
  publisher = {SAGE Publications Sage UK: London, England},
  year      = {1993},
}

\end{document}